\documentclass[a4paper,fleqn,usenatbib]{mnras}
\usepackage{color}
\usepackage{graphicx}
\usepackage{rotating}
\usepackage{lscape}
\usepackage{amsmath,amssymb,amstext}
\usepackage{multirow}
\usepackage{graphicx}
\usepackage{hyperref}

\title[Distance, reddening and three dimensional structure of the SMC]
{Distance, reddening and three dimensional structure of the SMC $-{\rm I}$: 
Using RRab stars}
\author[Sukanta Deb]{Sukanta Deb$^{1,2}$\thanks{E-mail: sukantodeb@gmail.com,~sukanta.deb@cottonuniversity.ac.in}\\
$^{1}$Department of Physics, Cotton University, Panbazar, Guwahati 781001,
Assam, India \\
$^{2}$Space and Astronomy Research Center, Cotton University, Panbazar, Guwahati 781001, Assam, India
}
\begin{document}

\date{Received on ; Accepted on }

\pagerange{\pageref{firstpage}--\pageref{lastpage}} \pubyear{2013}

\maketitle

\label{firstpage}

\begin{abstract}
We present a study of simultaneous determination of mean distance and reddening
 to the Small Magellanic Cloud (SMC) using the two photometric band RR Lyrae 
data. Currently available largest number of highly accurate and 
precise light curve data of the fundamental mode RR Lyrae stars (RRab) with 
better areal coverage released by the Optical Gravitational Lensing Experiment 
(OGLE)-IV project observed in the two photometric bands $(V,I)$ were utilised 
simultaneously in order to determine true distance and reddening independently 
for each of the individual RRab stars. Different empirical and theoretical 
calibrations leading to the determination of absolute magnitudes of RRab stars 
in the two bands, $V$ and $I$ along with  their mean magnitudes were utilised 
to calculate the apparent distance moduli of each of these RRab stars in these 
two bands. Decomposing the apparent distance moduli into true distance modulus 
and reddening in each of these two bands, individual RRab distance and 
reddening were estimated solving the two apparent distance moduli equations. 
Modeling the observed distributions of the true distance 
moduli and reddenings of the SMC RRab stars as Gaussian, the true mean 
distance modulus and mean reddening value  to the SMC  were found to be {\bf  
$\mu_{0}=18.909\pm0.148$ mag and $E(B-V)=0.066\pm0.036$ mag, respectively. This 
corresponds to a distance of $D = 60.506\pm 4.126$~kpc to the SMC. The three 
dimensional distribution of the SMC RRab stars was approximated as ellipsoid. 
Then using the principal axes transformation method \citep{deb14} we find the 
axes ratios of the SMC: $1.000\pm0.001,1.113\pm 0.002, 2.986\pm0.023$ with 
$i=3^{\circ}.156\pm0^{\circ}.188$ and $\theta_{\text{lon}}=38^{\circ}.027\pm0.577$.} These results are in agreement with other recent independent previous 
studies using different tracers and methodologies.           
\end{abstract}

\section{Introduction}
Studies of variable stars are crucial for addressing several key astrophysical
problems and issues. For example, distance determination is one of the most 
difficult but essential ingredients in modern astronomy. {\bf The pulsating stars 
such as Cepheids and RR Lyraes play pivotal roles in the definition of the 
distance ladder.} RR Lyrae stars provide a very useful means to obtain highly 
precise ($1-2\%$ error) distance measurements in the range $1-100$~kpc. They 
can be used to anchor Galactic and extra-galactic distances, to constrain the 
value of value of the Hubble constant $(H_{0})$ upto $3\%$ accuracy and to 
study the galactic structures \citep{deb14,deb15,beat16}. Furthermore using 
the light curves of these stars, it is also possible to determine their various
physical and chemical parameters including the reddening towards the line of 
sight \citep{deb10,kais17}. The population  II  distance scale 
relies on the absolute magnitudes of the RR Lyraes which in turn depend on 
the information about metallicity. Using RR Lyraes found in globular clusters, an estimate of the age of the cluster and 
hence a lower bound on the age of the Universe can be determined. Several 
empirical and theoretical relations connecting the metallicity and 
absolute  magnitudes for RR Lyrae stars available in multiwavelength bands 
provide the robust means of obtaining much improved distances and reddening 
estimations \citep{bono03, cate04,del05,marc15}.                      
           
Recently there has been a lot of interest in building a distance ladder 
based on Pop II stars such as RR Lyrae stars because of 
numerous advantages of these old stellar population and tracers of low-mass
stars, against the Pop I stars such as Cepheids. For instance, their presence 
in all kinds of galaxies as well in low density stellar haloes which have low 
reddening and relatively uniform and low metallicity populations make them  
robust and excellent standard candles independent of Cepheids. They are also 
much more numerous than the Population II or Type II Cepheids \citep{beat16}.  

The SMC is  a dwarf irregular galaxy connected by a hydrogen gas and 
stellar bridge to the Large Magellanic Cloud (LMC). Being the 
satellite galaxies of the Milky Way, they are useful for calibrations for 
many standard candles  \citep{grac14}. A large number of studies involving the
distance, reddening and three dimensional structure of the SMC rely on 
the available existing light curve data of variable stars generated from 
various astronomy missions and all sky surveys. Our present knowledge is 
limited by the flow of available data. During the last few decades, the OGLE 
project has revolutionized in the collection of a huge 
number of variable star light curve data of the Galaxy, the SMC and LMC in 
$(V,I)$-bands which was never done ever before. This revolution in 
collecting data by OGLE has led to the discovery of a number of variable stars
in the the Galaxy as well as Magellanic Clouds  and also helped to study a 
particular galaxy from its tracers of various standard candles. A large number 
of studies were devoted  to study the SMC using the OGLE archival data of RR 
Lyrae  stars and produced some significant results in recent decades, viz., \citet{deb10,kapa11,hasc11,hasc12,subr12,deb15}, to mention a few.  

The distance to the SMC is of vital interest to anchor and calibrate the 
extragalactic distance scale. There exist several studies in the literature 
which attempt to obtain the distance to the SMC using RRab, Cepheid and 
eclipsing binary light curve data ranging from optical to near-infrared (NIR) 
bands \citep[among others]{szew09,inno13,grac14,scow16}. Determinations of 
mean distance, reddening as well as three dimensional structure of the SMC 
utilising simultaneous $V$- and $I$-band data of RRab stars from 
the OGLE-IV database constitute the broader context of the present study. The 
availability of light curve data of an RRab star observed in two photometric 
$(V,I)$-bands has two major implications: (1) independent distance 
determination free from the effect of reddening and (2) reddening estimation. 
This is because the effect due to the reddening in the distance determination 
can be disentangled from the observed apparent distance modulus by using the 
light curve data of the same RRab star available in two photometric 
$(V,I)$-bands. Distance determination to the SMC or any other galaxy/globular 
cluster using RRab stars with single band photometric data require reddening 
estimations to be taken from other sources of reddening maps such as 
\citet{burn82,schl98,zari02} or some other reddening maps. For many targets 
where the foreground reddening is high and spatially variable, the reddening 
values can be directly determined from  the RRab stars themselves rather than 
relying on those aforementioned reddening maps if the data are available in 
two photometric bands. 

{\bf Although the method developed in the present study has been applied for 
the RRab stars, it is in general applicable to any periodic variable stars 
such as Cepheids which act as `standard candles'. This method is expected to 
give as accurate a result in the distance determination as those of the 
distance determination techniques based on Period-Wesenheit (PW) relations. 
This is because of the 
fact that the distance determinations in the present study as well as those 
based on PW relations are reddening-free by constructions. Despite giving 
almost the same result of distance determinations, this method is expected to 
provide an additional advantage over the PW-based methods as it also yields 
information about the reddening value towards the location of the star apart 
from finding its distance simultaneously. This information will serve to very 
useful in the determination of three dimensional structure as well as 
constructing the reddening map of the host galaxy/globular cluster.

Wesenheit functions are widely used in classical Cepheids to obtain 
reddening-free PW relations for accurate distance determinations when the data 
in two photometric bands are available \citep{mado82,free01,maja11}. These 
relations can also be adopted for RR Lyrae stars 
\citep[among others]{kova01,cris04,maja10,maja11,brag15,marc15}. In the case 
of RR Lyrae stars, the existence of reddening-free PW 
relations has been firmly supported from the findings of \citet{marc15} and 
\citet{brag15}, respectively. Using the PW relations for RR Lyrae stars, 
\citet{mart15} found the reddening-free distance modulus to the Sculptor dSph, 
\citet{brag15} derived the distance to the globular cluster M4 (NGC 6121), 
etc., among others. The distance determination using a different method, i.e. 
the Wesenheit function, which has been largely applied recently, would 
provide an interesting and independent alternative result to the present 
analysis.}

The main rationale of this paper is to  simultaneously exploit the available 
$V$- and $I$-band OGLE-IV data of more than $3000$ SMC RRab stars in order to 
independently determine their distances and reddenings. This information of 
individual RRab star will be useful to determine the true mean distance, 
reddening and three dimensional structure of the SMC as well as to 
construct its reddening map. In an earlier study of the SMC, \citet{deb15} 
used the $V$- and $I$-band RRab light curve data separately from the  OGLE-III 
catalog to study the three dimensional structure and metallicity distribution 
of the SMC using a very limited number of RRab stars available in the 
database. Apart from that, the study by \citet{deb15} does not use both the 
$V$- and $I$-band data simultaneously to determine the reddening values but 
uses the reddening values from the \citet{zari02} reddening map. 

The paper is organised as follows: {\bf Section~\ref{data} describes the OGLE-IV 
RRab data of the SMC and sample selection for the analysis in the present 
study. The methodologies developed here for the distance determination and 
reddening estimations are described in Section~\ref{methodology}. Metallicity 
relation of \citet{neme13} was used to derive $M_{V}$ and $M_{I}$ values of 
the sample of RRab stars. Section \ref{selection} demonstrates the application 
of the technique to an RRab star in the OGLE-IV database, while 
Section~\ref{comp_other} describes the comparison of the values of distance 
and reddening with those obtained using other empirical and theoretical 
relations. For comparison with the distance obtained in the present study, 
distances are calculated from reddening-free Wesenheit distance 
moduli ($\mu_{W}$) using the observed and absolute Wesenheit 
magnitudes. Absolute Wesenheit magnitudes of RRab stars are calculated using 
the theoretical Period-Wesehnehit-Metallicity (PWZ) relation of \citet{brag15}. 
Mean distance and reddening determination of the SMC are discussed in 
Section~\ref{distance}, whereas Section~\ref{error} gives an 
account of error estimation of the derived quantities. The period-colour 
relation derived using the independent reddening values in the present study 
is discussed in Section~\ref{pc}. Determination of the three 
dimensional structure of the SMC applying different fitting algorithms is 
discussed in Section~\ref{morphology}. Results obtained from the above 
analyses are compared with those obtained  using the \citet{smol05} metallicity 
relation and with other values available in the literature in 
Section~\ref{compare}. Lastly, the summary and conclusions of the present 
investigation are presented in Section~\ref{summary}.  }
\begin{figure*}
\vspace{0.02\linewidth}
\begin{tabular}{cc}
\vspace{+0.01\linewidth}
  \resizebox{0.40\linewidth}{!}{\includegraphics*{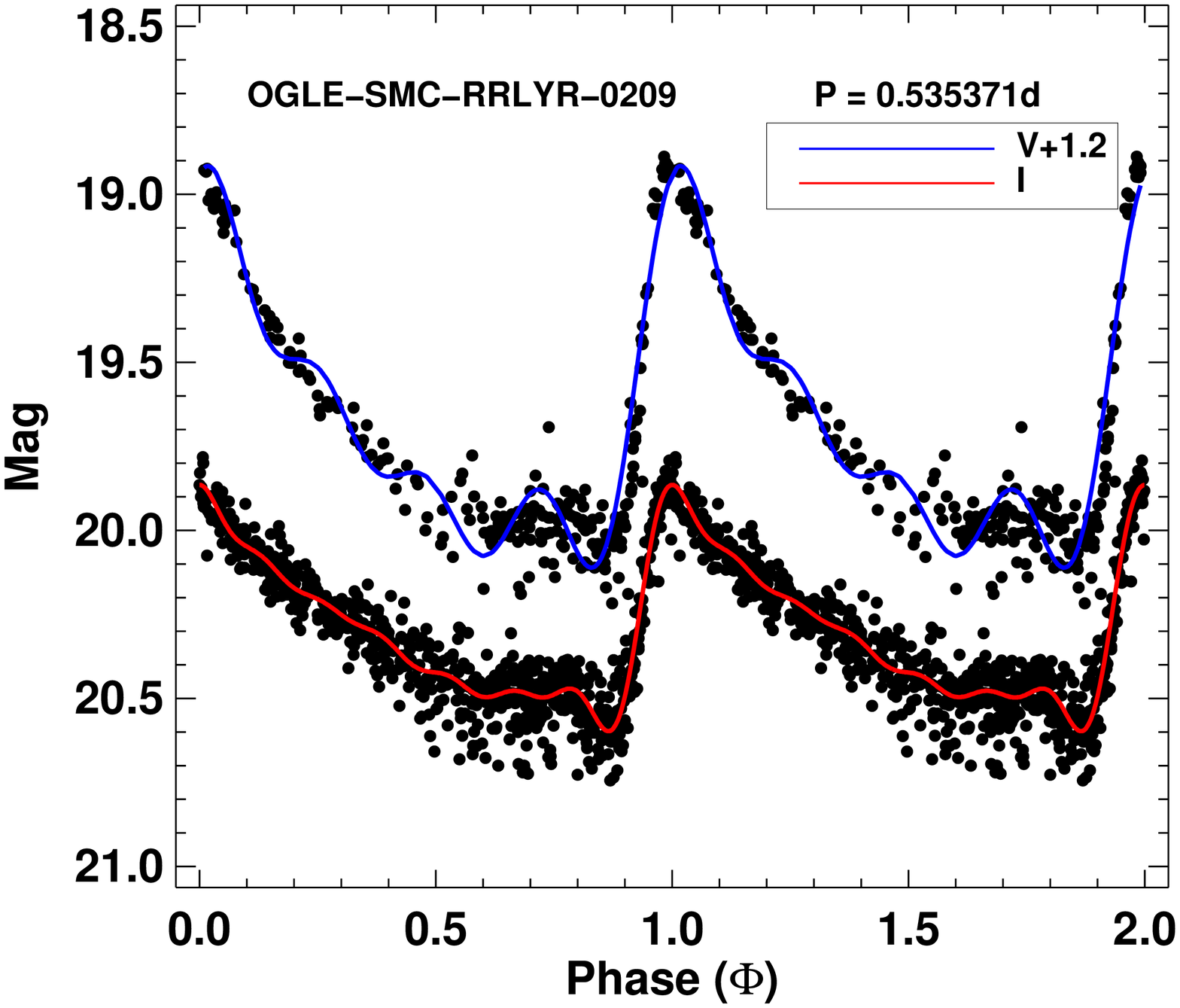}}&
  \resizebox{0.40\linewidth}{!}{\includegraphics*{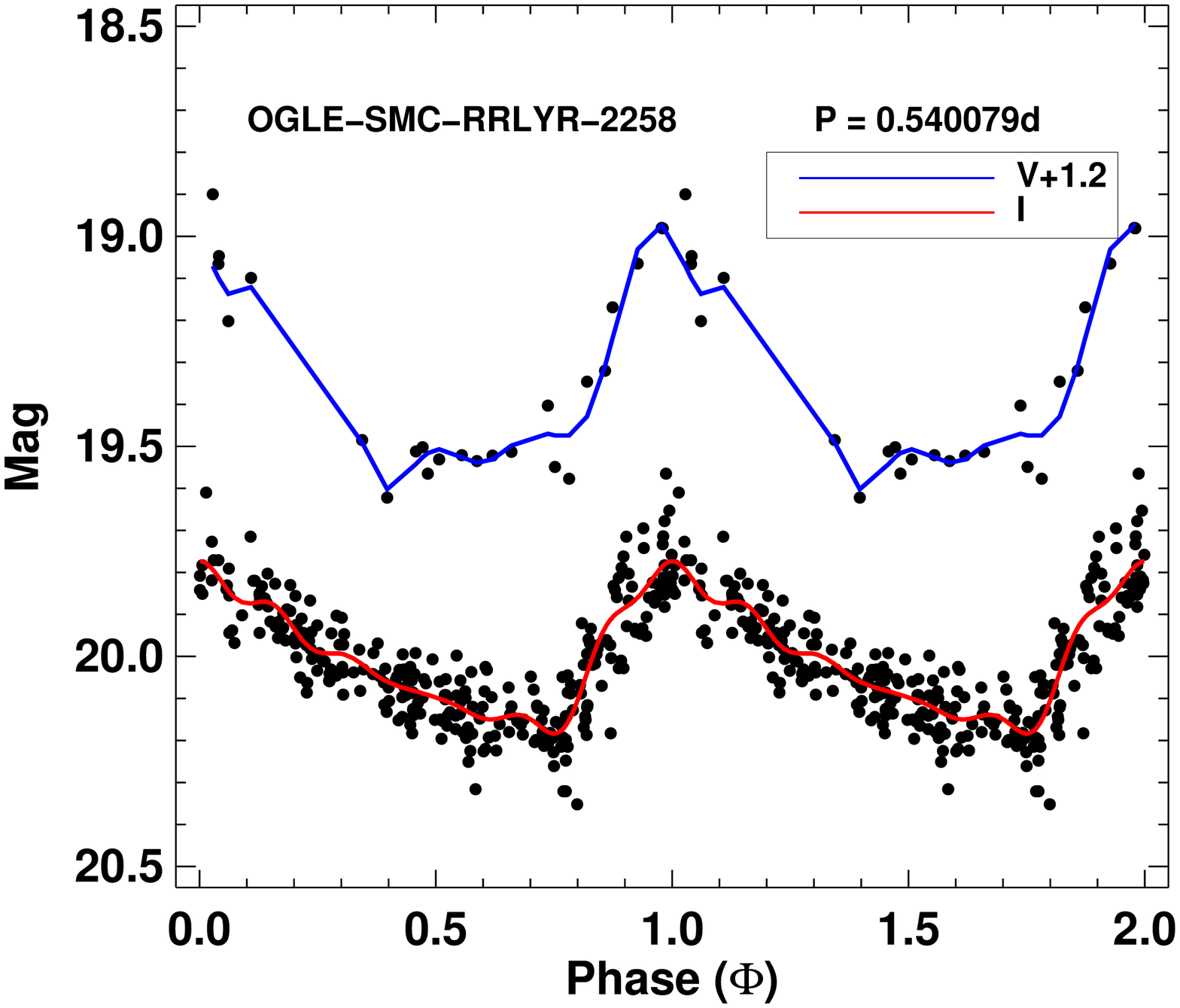}}\\
\vspace{0.01\linewidth}
\resizebox{0.40\linewidth}{!}{\includegraphics*{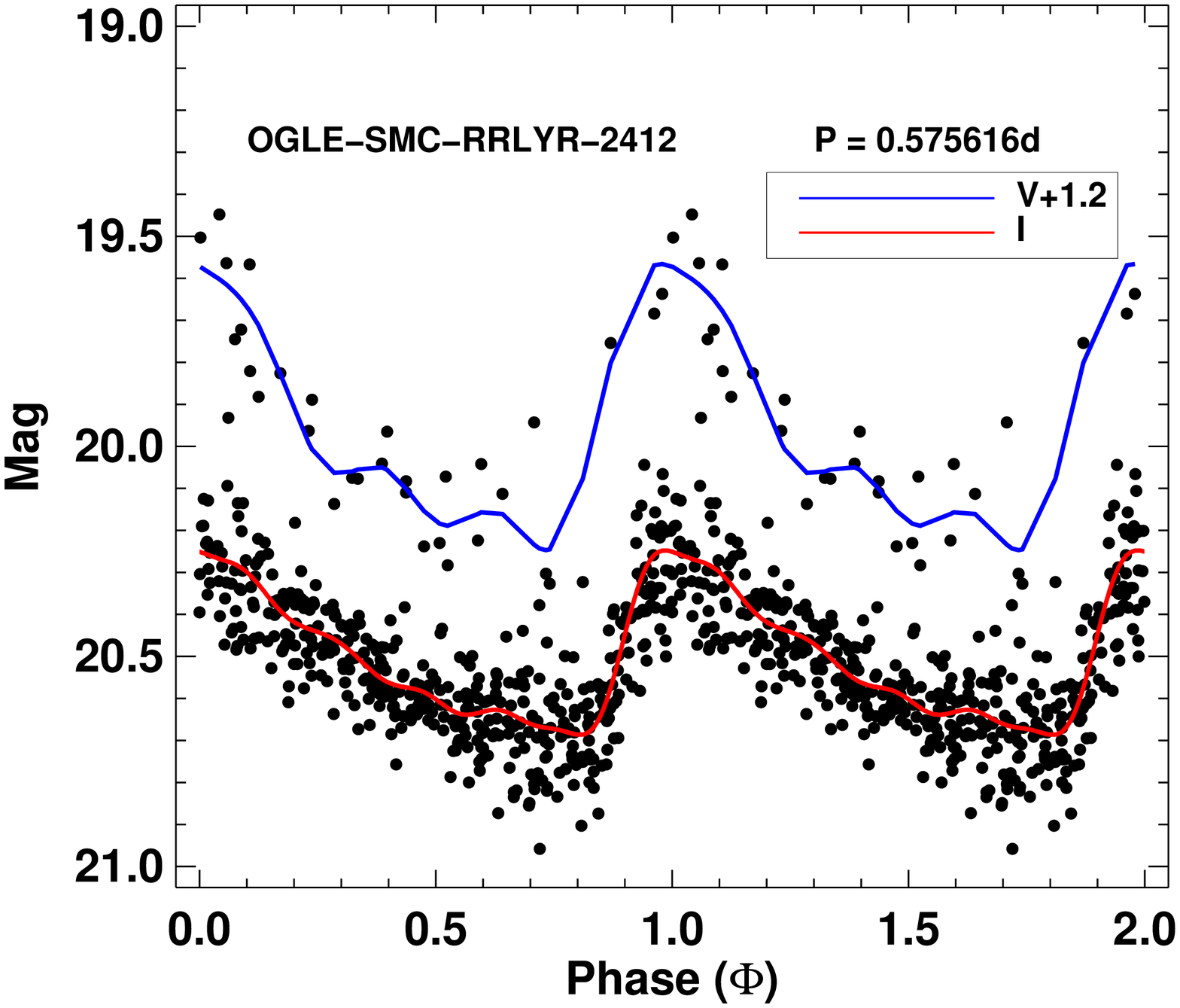}}&
\resizebox{0.40\linewidth}{!}{\includegraphics*{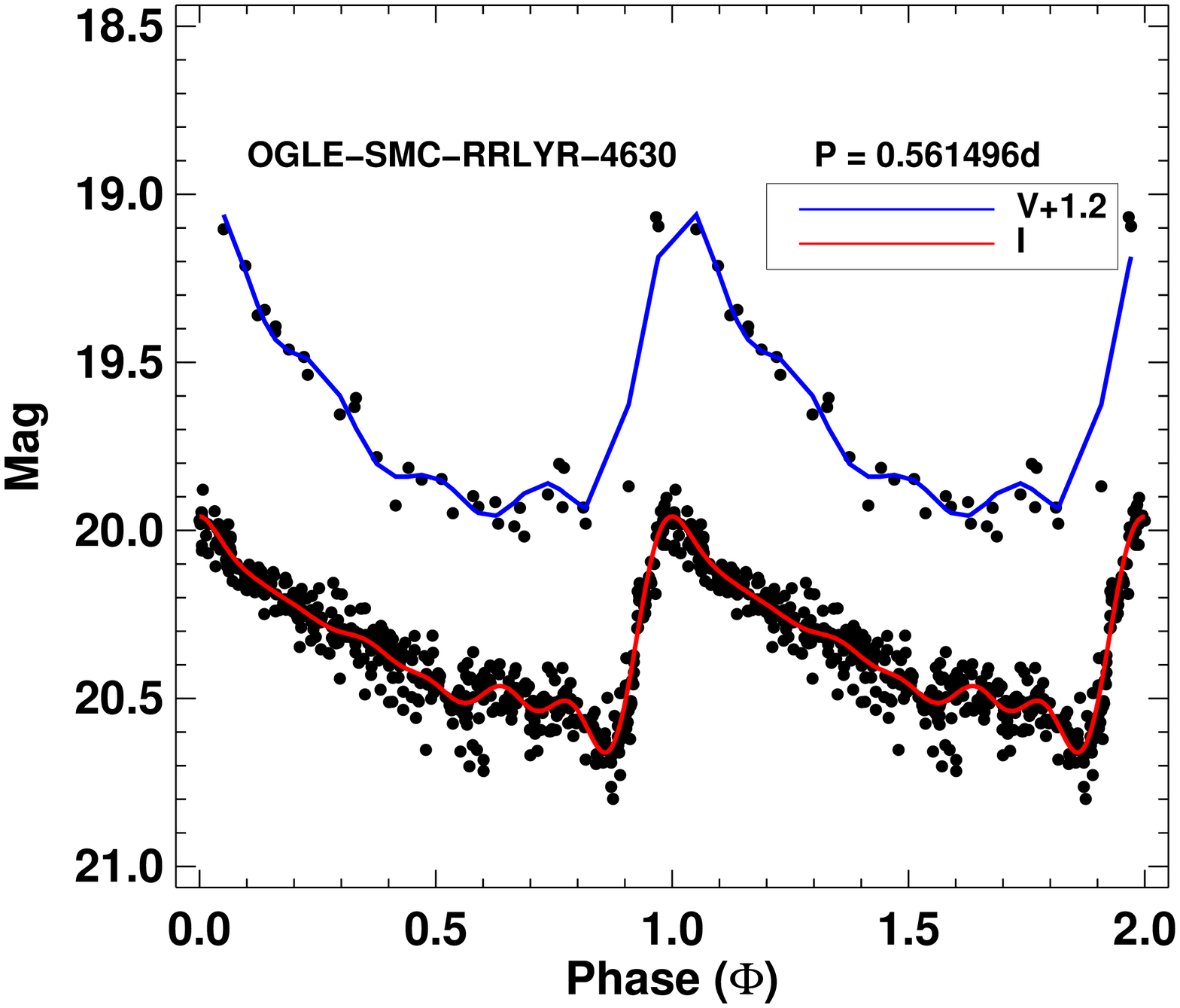}}\\
\vspace{-0.04\linewidth}
\end{tabular}
\caption{A collection of four randomly selected sample of RRab stars with 
their corresponding OGLE-IV IDs in the $V$ and $I$ bands. The Fourier fits
for the $V$ and $I$-band are shown in blue and red solid lines, respectively.} \label{lcplots}
\end{figure*}
\section{The Data and Sample Selection}
\label{data}
{\bf
The OGLE-IV is one of the largest sky variability surveys worldwide in recent
times which covers over $3000$ square degrees in the sky and regularly monitors 
over a billion sources. The main targets of this project include
the inner Galactic bulge and the Magellanic system. The photometric accuracy 
of this project is of the order of $0.001-0.005$ mag \citep{udal15,sosz16a}. 
The number of new sources in this part of the project has at least doubled as 
compared to the last OGLE-III project \citep{udal15}. The details of the 
OGLE-IV project, the instruments such as the telescope, mosaic CCD (Charge 
Coupled Device) camera and the $(V,I)$-filters are provided in \citet{udal15}. 

The OGLE-IV collection of RRab stars is an extension of OGLE-III catalog to 
the regions covered by the OGLE-IV fields. The areal coverage of the OGLE-III 
catalog was $54$ square degrees of the sky which covers mostly the central 
regions of the LMC and SMC.  The recently released OGLE-IV catalog covers 
around $650$ square degrees of the sky which include the larger part of the 
Magellanic system covering the outermost parts as well as the Magellanic 
Bridge connecting them. The OGLE-IV database contains the data obtained with 
the $1.3$-meter Warsaw telescope equipped with a $32$-detector mosaic
CCD camera located at Las Campanas Observatory (operated by the Carnegie 
Institution for Science), Chile.   
In order to carry out the present study, RRab stars were selected from the 
OGLE-IV catalog of variable stars that consist of $5$-year of archival data 
observed between Match $2010$ and July $2015$. The database contains Johnson 
$V$ and Cousins $I$ band light curve data, with the majority of the 
observations obtained in the $I$-band \citep{sosz16b}.

The classification of the SMC RRab stars in the OGLE-IV catalog was based on 
the periods, amplitudes and Fourier parameters of the light curves obtained 
from the Fourier cosine decomposition method \citep{sosz16b}. The catalog 
contains information about the OGLE ID, mean magnitudes of the light curves 
$\overline{m}_{I}, \overline{m}_{V}$, in the $I$ and $V$-bands, respectively, 
period $P$, error in the period $\sigma_{P}$, time of maximum brightness 
$t_{0}$, amplitude in the $I$-band $A_{I}$, Fourier parameters $R_{21}, 
\phi_{21}, R_{31}, \phi_{31}$ in the $I$-band as well as time-series 
photometry in the $(V,I)$-bands. The OGLE-IV catalog provides new $18158$ RR 
Lyrae stars in the LMC and SMC which were not present in the earlier 
phases of the OGLE survey. This is by far the largest number of RRab light 
curve data generated with complete phase coverage in the $V$- and $I$-bands 
obtained in the SMC. The OGLE-III catalog consists of $1933$ SMC RRab stars 
with their number increased to $4961$ in the new data release of OGLE-IV.  
The  $I$-band RRab light curve data  of OGLE-IV have even phase coverage 
containing an average of $400$ accurate and precise photometric observations 
with the exposure time of $100$~s. On the other hand, $V$-band light curves of 
RRab star contain only an average of $40$ data points ($10\%$ of the $I$-band 
observations) per light curve with the exposure time of $150$~s 
\citep{sosz16b}. Since the database does not provide standard errors in the 
mean magnitudes as well in the Fourier parameters, we do not use these values  
for the analysis done in this paper. We use only the light curve data 
available in both the two bands ($V,I$) along with the information about their 
periods ($P$) and epochs of maximum light ($t_{0}$).

The available $V$- and $I$-band photometric light curves of common SMC RRab 
stars from the OGLE-IV catalog were matched. The matched light curves were 
further subjected to pre-selection. The $I$-band light curves contain comparatively 
more number of data points than the corresponding $V$-band light curves. 
Out of $4961$ RRab stars available in the catalog, there are $4769$ RRab stars 
which have complementary light curve data available in both the $V$- and 
$I$-band. We have selected only those common light curves which contain 
at least $20$ data points in the $V$-band for reliable light curve parameter 
estimations. This condition further reduces the number of common RRab stars in 
both the two bands to $3931$ for analysis. The light curves of these $3931$ 
RRab stars were Fourier decomposed using sinusoidal law to derive their mean 
apparent magnitudes, amplitudes in the $I$ and $V$-band and Fourier parameters 
for the $I$ and $V$-band light curves, respectively. Apparent mean magnitudes 
and  Fourier parameters in the $I$ and $V$-band are obtained from the Fourier 
sine decomposition of the light curves \citep{deb15}
\begin{align}
\label{eq:fd}
m_{\lambda}(t)= \overline{m}_{\lambda}+\sum_{i=1}^{N}A_{i}\sin{\left[i\omega(t-t_{0})+\phi_{i}\right]},~~\lambda=\left(I,V\right)
\end{align}
using the seven and fourth order Fourier fits for the $I$ and $V$ bands, 
respectively. Here $\overline{m}_{\lambda}$ is the mean magnitude, 
$\omega=2\pi/P$ is the angular frequency and $t$ is the time of observations. 
$t_{0}$ represents the epoch of maximum light. The phased light curves 
$m_{\lambda}(\Phi)$ are obtained using 
\begin{align*}
\Phi=\frac{\left(t-t_{0}\right)}{P}-Int\left(\frac{t-t_{0}}{P}\right),
\end{align*}                  
where $\Phi \in [0,1]$ represents one pulsational cycle of the RRab stars. 
A collection of four randomly selected sample of RRab stars with thier 
corresponding OGLE IDs and periods in the $V$ and $I$-band is shown in Fig.~\ref{lcplots}.  The corresponding Fourier fits for the respective $V$ and 
$I$-band are also shown in blue and solid colour solid lines. 
The phase differences $\phi_{i1}=\phi_{i}-i\phi_{1}$ and amplitude ratios
$R_{i1}=\left(A_{i}/A_{1}\right), i> 1$ are evaluated and standard errors are
determined following \citet{deb10} for both the $I$ and $V$ bands, 
respectively. It should be noted that $\phi_{i1} \in [0,2\pi]$ radian. Since 
the $I$-band light curves contain more number of data points the Fourier 
parameters obtained in this band will be more accurate and precise as compared 
to the corresponding $V$-band Fourier parameters. In this paper we use 
$\phi_{31}^{I}$ to denote the Fourier phase parameter $\phi_{31}$ in the $I$ 
band obtained from the Fourier sine decomposition. The light curve parameters 
obtained from the Fourier sine decomposition method as given by 
equation~(\ref{eq:fd}) are listed in Table~\ref{four1}.

A comparison of the light curve parameters of $3931$ stars in the present 
study (along abscissa) with those available from the OGLE-IV database 
\citet{sosz16b} (along ordinate) in shown in Fig.~\ref{pap_lit} as 
scatter plots. As shown in the figure $\phi_{31}({\rm cosine})$ (This work) 
has been calculated by subtracting $\pi$  from $\phi_{31}^{I}$ (provided in the 
last column of Table~\ref{four1}) obtained from the Fourier sine decomposition 
in the present study just for comparison with the value of $\phi_{31}$ 
as given in the database which is obtained from the Fourier cosine 
decomposition. The sine and cosine $\phi_{31}$ phase parameters differ by 
$\pi$ radians, i.e, $\phi_{31}(\text{cosine})=\phi_{31}(\text{sine})-\pi$ 
\citep{deb10,neme11}.  
If the value of $\phi_{31}({\rm cosine})$ (This work) comes out to be 
negative, then a value of $2\pi$ is added to it so that its value always lies 
in the interval $[0,2\pi]$. 

The Fourier phase parameter $\phi_{31}$ is often used to determine the 
metallicity of an RRab star along with the information of the period $P$ of an 
RRab star \citep[among others]{jurc96,smol05,neme13}. A histogram plot of the distribution 
of standard errors in $\phi_{31}^{I}$ is shown in Fig.~\ref{ephi31}, where $\sigma_{\phi_{31}^{I}} <0.5$. In order to select a clean sample of the SMC RRab 
stars for the present analysis, we apply the selection criteria based on 
OGLE-determined periods $P$, the mean magnitudes 
$(\overline{m}_{V},\overline{m}_{I})$, observed colours $(V-I)$, 
amplitudes ($A_{V},A_{I}$), error in the $I$-band Fourier phase parameter 
$\sigma_{\phi_{31}}$ determined from the Fourier analysis of the light curves 
as described by equation~(\ref{eq:fd}) and metallicities $[Fe/H]$. RRab stars 
with $P\ge 0.4$~d, $\overline{m}_{V} \ge 18.5$ mag, $\overline{m}_{I} \ge 18.0$ mag, $0.4 \le (V-I) \le 0.8$ mag, $0.2\le A_{V} \le 1.5 $ mag, 
$0.1 \le A_{I} \le 1.2$ mag, $\sigma_{\phi_{31}^{I}}<0.5$, 
$-2.70 < [Fe/H] < 0$ dex were chosen for the analysis. Here $[Fe/H]$ 
represents the metallicity in the \citet{zinn84} scale. Most of the selection 
criteria were adapted from \citet{deb15} while a few of them, viz., 
$\sigma_{\phi_{31}}$ from  \citet{deb10}; colour from \citet{hasc12} with 
slight modifications.  Certain selection criteria were always applied in the 
literature in order to choose a clean sample of RRab stars belonging to a 
particular galaxy free from any possible contamination due to the foreground 
objects of any other galaxy \citep{pejc09,hasc11,deb15}. The application of 
this final selection criteria applied on the $3931$ RRab stars further reduces 
their number to $3522$ for the light curve analysis.     
\begin{figure*}
\vspace{0.02\linewidth}
\begin{tabular}{ccc}
\vspace{+0.01\linewidth}
  \resizebox{0.32\linewidth}{!}{\includegraphics*{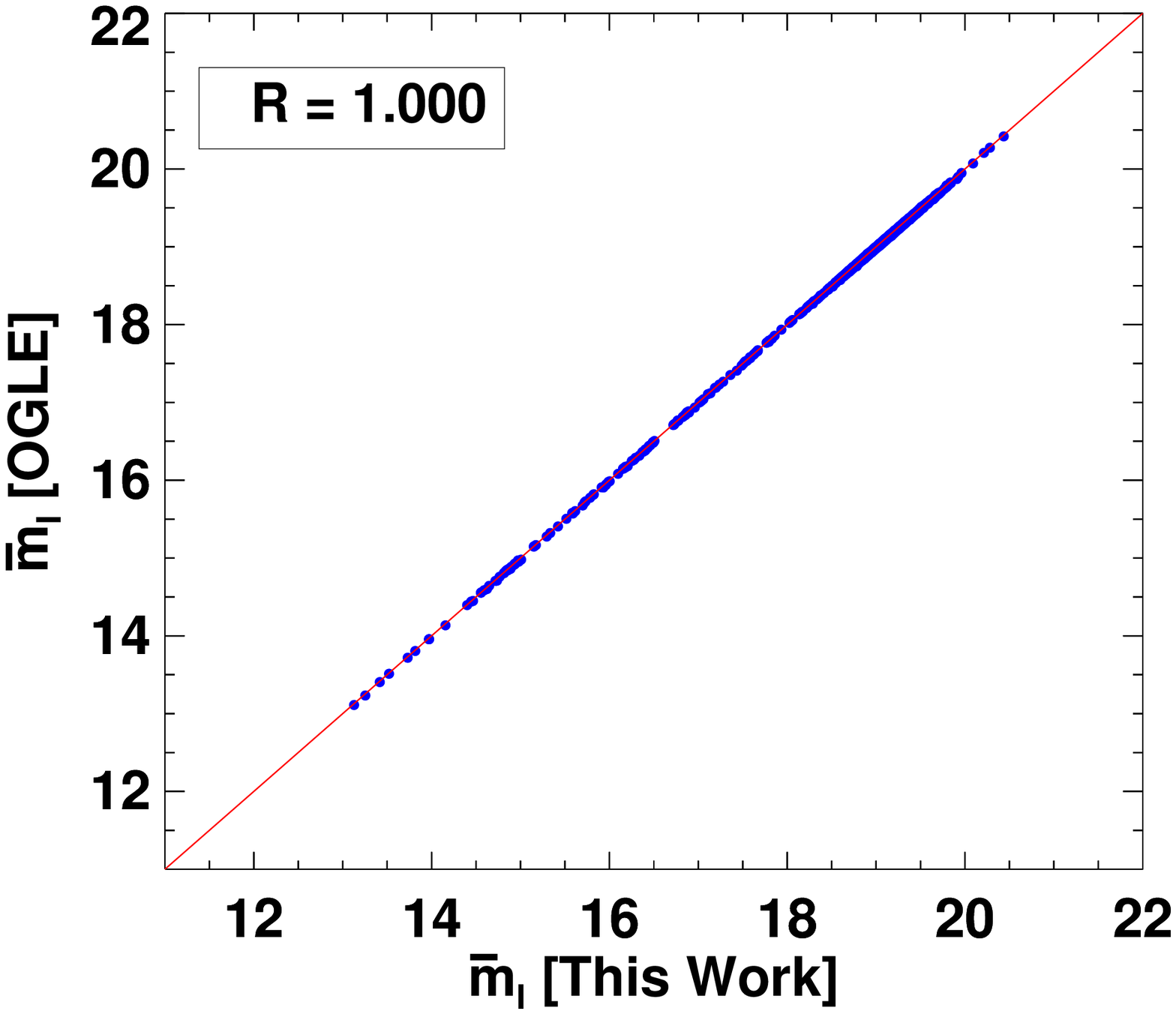}}&
    \resizebox{0.32\linewidth}{!}{\includegraphics*{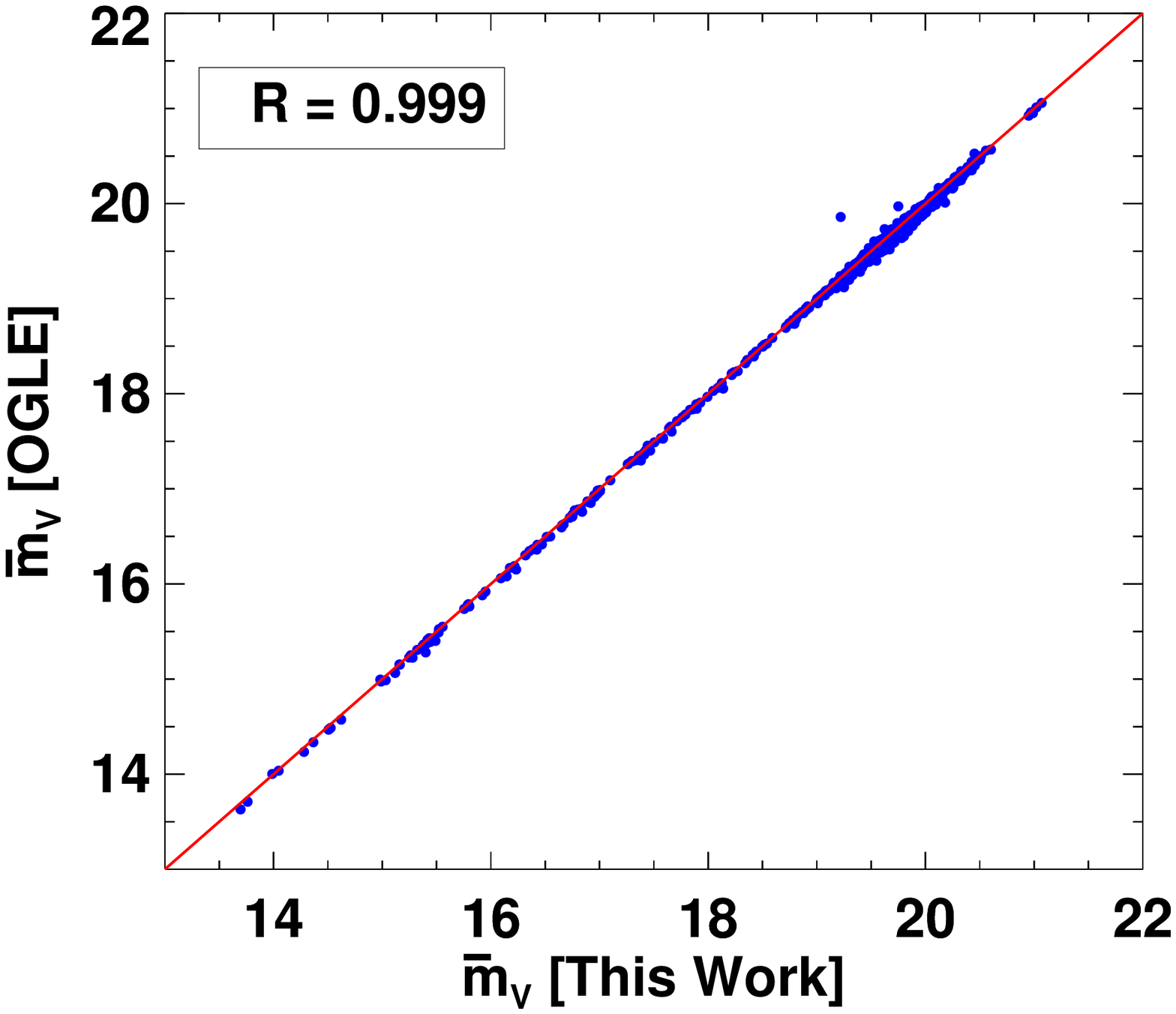}} &
  \resizebox{0.32\linewidth}{!}{\includegraphics*{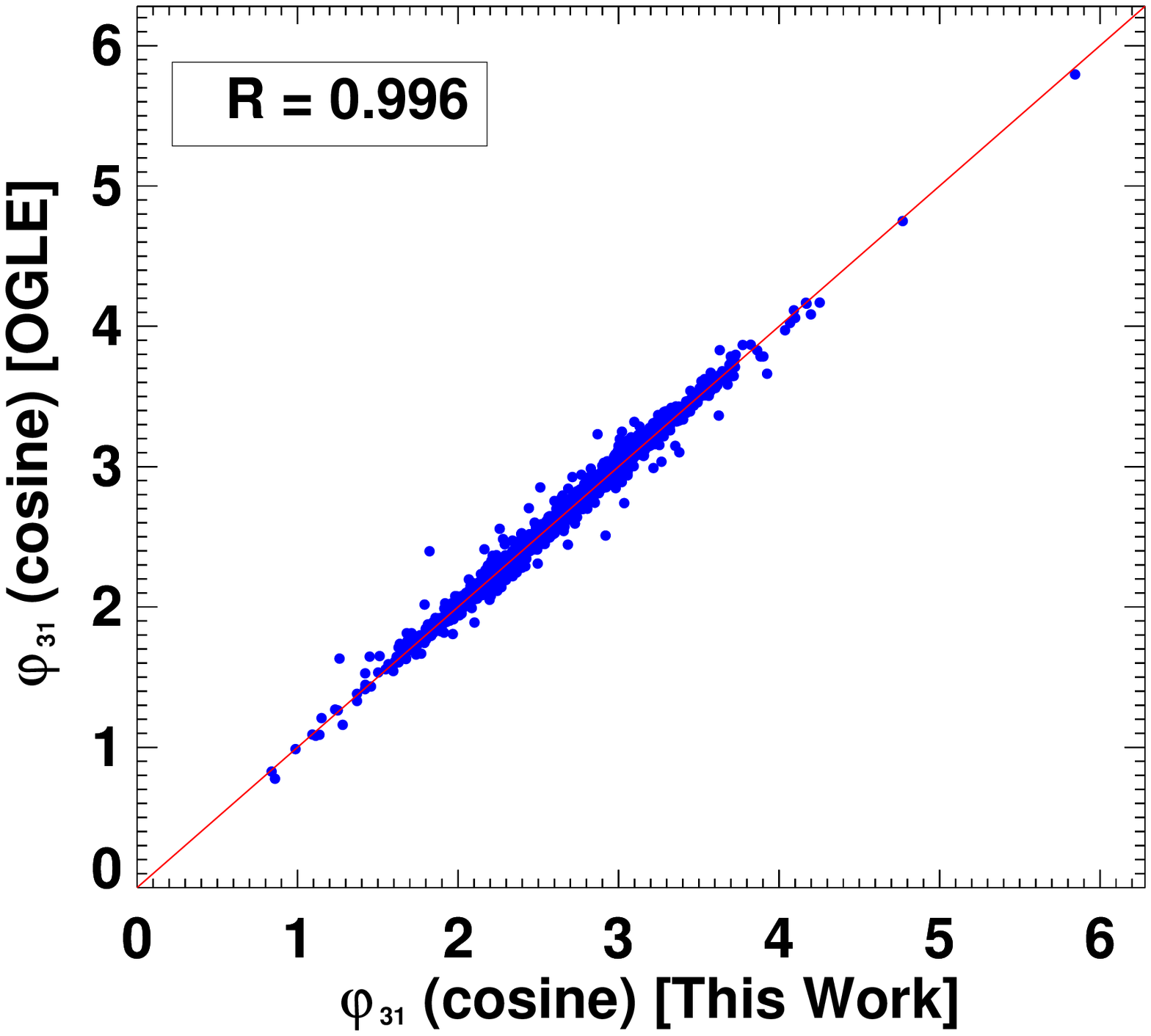}}\\
\vspace{-0.04\linewidth}
\end{tabular}
\caption{{\bf Scatter plots of the light curve parameters of $3931$ stars 
determined in the present study (along abscissa) with those available from 
\citet{sosz16b} (along ordinate). The clustering of data points along a line 
of zero intercept and unit slope in each of the plots indicates a high value 
of linear positive correlation coefficient $R$. $\phi_{31}({\rm cosine})$ 
(This work) has been calculated by subtracting $\pi$ radians from 
$\phi_{31}^{I}$ obtained from the Fourier sine decomposition in the present 
study just for comparison with \citet{sosz16b}. If the value of 
$\phi_{31}({\rm cosine})$ (This work) comes out to be negative, then a value 
of $2\pi$ is added to it so that its value lies in the interval $[0,2\pi]$. It 
should be noted that the value of $\phi_{31}$ given in the OGLE-IV SMC RRab 
database \citep{sosz16b} has been obtained from the Fourier cosine 
decomposition in the $I$-band.}}
\label{pap_lit}
\end{figure*}

Availability of a large number of SMC RRab light curve data in OGLE-IV database 
with complete phase coverage  in two different photometric bands $(V,I)$ with 
an improved areal coverage provides a unique opportunity to determine the 
distance and reddening of each of the individual stars. Since the database of 
RRab stars is substantially larger than any previously studied subset of 
SMC RRab stars, this wealth of new data can in turn be used to refine our 
knowledge about the distance scale and reddening distribution to the SMC. This 
will lead to a detailed understanding of the three dimensional structure, 
dust  and metallicity distribution of the SMC.  }
\begin{table*}
\centering
\caption{{\bf A sample of the light curve parameters of $3931$ OGLE-IV SMC RRab 
stars used for the analysis obtained from the Fourier sine decomposition of the 
light curves as given by equation~(\ref{eq:fd}). The full table is available 
as supplementary material in the online version of this paper.}}
\label{four1}
\scalebox{1}{
\begin{tabular}{ccccccccccc}
\\
\hline
\hline
OGLE ID & $P$[days] & $\overline{m}_{I}$ ~[mag]& $\overline{m}_{V}$ ~[mag] & $A_{I}$~ [mag] & $A_{V}$~ [mag] & $\phi_{31}^{I}$ ~[rad] \\
&&$\sigma \overline{m}_{I} $ [mag]&$\sigma \overline{m}_{V}$ [mag]  & $\sigma A_{I}$ [mag] & $\sigma A_{V}$ [mag] & $\sigma \phi_{31}^{I}$ [rad] \\\hline
OGLE-SMC-RRLYR-0001& 0.5588145&    19.067&    19.584&     0.674&     0.988&     5.360\\
&&     0.002&     0.005&     0.026&     0.015&     0.060\\
OGLE-SMC-RRLYR-0002& 0.5947940&    19.011&    19.613&     0.421&     0.695&     5.944\\
&&     0.002&     0.005&     0.055&     0.192&     0.117\\
OGLE-SMC-RRLYR-0003& 0.6506795&    19.158&    19.772&     0.276&     0.399&     6.233\\
&&     0.002&     0.002&     0.126&     0.075&     0.433\\
OGLE-SMC-RRLYR-0005& 0.5652651&    19.056&    19.638&     0.486&     0.638&     5.351\\
&&     0.002&     0.005&     0.111&     0.359&     0.099\\
OGLE-SMC-RRLYR-0006& 0.5471843&    19.013&    19.585&     0.708&     1.033&     5.259\\
&&     0.002&     0.005&     0.123&     0.105&     0.060\\
OGLE-SMC-RRLYR-0008& 0.6328767&    19.156&    19.742&     0.350&     0.551&     6.203\\
&&     0.002&     0.002&     0.070&     0.097&     0.154\\
OGLE-SMC-RRLYR-0010& 0.5530273&    19.260&    19.792&     0.377&     0.727&     5.690\\
&&     0.002&     0.006&     0.029&     0.064&     0.128\\
OGLE-SMC-RRLYR-0011& 0.5957643&    19.170&    19.803&     0.450&     0.765&     5.740\\
&&     0.002&     0.003&     0.227&     0.219&     0.100\\
OGLE-SMC-RRLYR-0012& 0.6256321&    19.195&    19.803&     0.454&     0.676&     6.038\\
&&     0.002&     0.002&     0.098&     0.059&     0.108\\
OGLE-SMC-RRLYR-0013& 0.6711785&    19.028&    19.621&     0.354&     0.470&     6.132\\
&&     0.002&     0.007&     0.206&     0.117&     0.155\\ \hline
\end{tabular}
}
\end{table*}
\section{Methodology}
\label{methodology}
{\bf 
We know that the apparent distance modulus ($\mu_{\lambda}$) 
observed in a particular wavelength band $(\lambda)$ can be decomposed into a 
sum of true distance modulus ($\mu_{0}$) and interstellar extinction 
($\mathcal{A}_{\lambda}$) as follows \citep{free01,kanb03}:
\begin{align}
\label{eq:mulambda}
\mu_{\lambda}=& \mu_{0}+\mathcal{A}_{\lambda} \nonumber \\
\Rightarrow \mu_{\lambda}=& \mu_{0}+R_{\lambda}E(B-V), 
\end{align}      
where $R_{\lambda}$ is the ratio of the total to selective absorption in a 
particular wavelength band ($\lambda$) and is defined as 
\begin{align}
R_{\lambda}=& \frac{\mathcal{A}_{\lambda}}{E(B-V)}. 
\end{align}
$R_{\lambda}$ is to be taken from a given reddening law and has to be held 
fixed. $E(B-V)$ is the interstellar reddening along the line of sight. 
$\mu_{\lambda}$ is defined as \citep{carr06}:
\begin{align}
\label{eq:mu}
\mu_{\lambda}=& \overline{m_{\lambda}}-M_{\lambda},  
\end{align}
where $\overline{m}_{\lambda}, M_{\lambda}$ denote the apparent mean  and absolute 
magnitudes of the star in the particular wavelength band $\lambda$, 
respectively. $\overline{m}_{\lambda}$ can be estimated from the Fourier 
decomposition of the light curve and the determination of $M_{\lambda}$ relies 
on various empirical and theoretical calibrations of  RRab stars.

Since we have the light curve data of the RRab stars in two bands, viz. $V$ 
and $I$, we can write down the equation~(\ref{eq:mulambda}) as \citep{kanb03}:
\begin{align}
\label{eq:muiv}
\mu_{I}=& \mu_{0}+R_{I}E(B-V) \nonumber \\
\mu_{V}=& \mu_{0}+R_{V}E(B-V).  
\end{align}
The above linear system of two equations contain two variables, viz. $\mu_{0}$
and $E(B-V)$ and hence can be solved exactly. The solutions of the above two
equations will yield individual values of reddening and true distance modulus
as follows:
\begin{align}
\label{simul}
E(B-V)=& \frac{\mu_{V}-\mu_{I}}{R_{V}-R_{I}} \\
\mu_{0}=&\mu_{I}-R_{I}E(B-V)~~\nonumber \\ 
&\text{or}~~\nonumber \\
\mu_{0}=&\mu_{V}-R_{V}E(B-V).~~\nonumber
\end{align}
Distance can be calculated using the following relation \citep{carr06}:
\begin{align}
\label{eq:distance}
D [\rm pc] = 10^{\frac{\left(\mu_{0}+5\right)}{5}}.
\end{align}
Fixing the reddening law \citep{card89} and assuming $R_{V}=3.23$, one can 
obtain $\frac{\mathcal{A}_{I}}{\mathcal{A}_{V}}=0.61$ \citep{inno13}. Using 
these values, we find out the relation between $E(V-I)$ and $E(B-V)$ as well 
as determine the value of $R_{I}$ as follows. We know that $R_{V}$ is given as  
\begin{align*}
R_{V}=& \frac{\mathcal{A}_{V}}{E(B-V)},
\end{align*} 
where $E(B-V)$ is defined by 
\begin{align*}
E(B-V)=\mathcal{A}_{B}-\mathcal{A}_{V}.
\end{align*}
Here $\mathcal{A}_{B}$ and $\mathcal{A}_{V}$ denote the interstellar 
extinctions in the $B$ and $V$ bands, respectively. Firstly let us try to find 
a relation between $E(V-I)$ and $E(B-V)$. We know that
\begin{align}
\label{av1}
E(V-I)=&\mathcal{A}_{V}-\mathcal{A}_{I}\nonumber \\
\Rightarrow E(V-I)=& \mathcal{A}_{V}\left(1-\frac{\mathcal{A}_{I}}{\mathcal{A}_{V}}\right) \nonumber \\
\Rightarrow E(V-I)=& \mathcal{A}_{V}\left(1-0.61\right) \nonumber \\
\Rightarrow E(V-I)=& \mathcal{A}_{V}\left(0.390\right) \nonumber\\
\Rightarrow \mathcal{A}_{V} =&\frac{1}{0.390}E(V-I) \nonumber \\
\Rightarrow \mathcal{A}_{V} =& 2.564E(V-I) 
\end{align}
Also, we have
\begin{align}
\label{av2}
\mathcal{A}_{V}=&R_{V}E(B-V) \nonumber\\
\Rightarrow \mathcal{A}_{V}=& 3.23E(B-V)   
\end{align} 
Comparing equations (\ref{av1}) and (\ref{av2}), we get 
\begin{align*}
2.564E(V-I)=& 3.23E(B-V) \\
\Rightarrow E(V-I)=& \frac{3.230}{2.564}E(B-V) \\
\Rightarrow E(V-I)=&1.26 E(B-V).    
\end{align*}
This relation is nearly identical to $E(V-I)=1.265E(B-V)$ obtained by 
\cite{inno16}. Now I calculate the value of $R_{I}$ as follows. We know that 
\begin{align*}
\mathcal{A}_{I}=& R_{I}E(B-V) \\
\Rightarrow   \frac{1}{E(B-V)} = & \frac{R_{I}}{\mathcal{A}_{I}}  \\
\Rightarrow \frac{\mathcal{A}_{V}}{E(B-V)}=& \frac{\mathcal{A}_{V}}{\mathcal{A}_{I}}R_{I}~~\left(\text{Multiplying both  sides by~}\mathcal{A}_{V}\right) \\
\Rightarrow R_{V}=& \frac{1}{0.61}R_{I} \\
\Rightarrow R_{I}=&0.61\times R_{V} \\
\Rightarrow R_{I}=&0.61\times 3.23 \\
\Rightarrow R_{I}=&1.97.      
\end{align*}
The values of $R_{V}$ and $R_{I}$ were held fixed in the above linear system 
of equations~(\ref{eq:muiv}). 

We  now show that the relations given by equation~(\ref{eq:muiv}) can be 
reduced to a form of reddening-free Wesenheit function \citep{free01}. We have 
found that the true distance modulus can be written as 
\begin{align}
\label{eq:free01_1}
\mu_{0}=& \mu_{I}-R_{I}E(B-V) \nonumber \\
\Rightarrow \mu_{0}= &\mu_{I}-R_{I}\left(\frac{\mu_{V}-\mu_{I}}{R_{V}-R_{I}}\right) \nonumber \\
\Rightarrow \mu_{0}= & \mu_{I}-\left(\frac{R_{I}}{R_{V}-R_{I}}\right) \left(\mu_{V}-\mu_{I}\right) \nonumber \\
\Rightarrow \mu_{0}=& \mu_{I}-R^{I}_{VI} \left(\mu_{V}-\mu_{I}\right),
\end{align}
where $R^{I}_{VI}=\frac{R_{I}}{R_{V}-R_{I}}=\frac{\mathcal{A}_{I}}{E(V-I)}$. 
In the present case, the value of $R^{I}_{VI}$ is $1.563$. 
Similarly, by taking the $V$-band distance modulus relation, we can show that
\begin{align}
\label{eq:free01}
\mu_{0}=& \mu_{V}-\left(\frac{R_{V}}{R_{V}-R_{I}}\right)\left(\mu_{V}-\mu_{I}\right) \nonumber \\
\mu_{0}=& \mu_{V}-R^{V}_{VI} \left(\mu_{V}-\mu_{I}\right),
\end{align}
where $R^{V}_{VI}=\frac{R_{V}}{R_{V}-R_{I}}=\frac{\mathcal{A}_{V}}{E(V-I)}$. The value 
of $R^{V}_{VI}$ is $2.563$. The quantity $\mu_{0}$ is called the reddening-free 
distance modulus or the Wesenheit function  \citep{free01,kanb03}. The 
above procedure is equivalent to a reddening-free Wesenheit 
index $W=V-R(V-I)$ \citep{mado82,free01}. The relation given by 
equation~(\ref{eq:free01}) is exactly the same as derived by 
\citet{free01}. Equivalently, the reddening-free distance modulus can be 
determined with the data available in two photometric $(V,I)$-bands using 
Equation~(\ref{eq:free01_1}) or (\ref{eq:free01}). In order to determine the 
absolute magnitudes of the RRab stars in the two bands $(V,I)$, we use the 
following relations:
\begin{align}
\label{eq:absmag}
M_{V}=& 0.23[Fe/H]+0.984 \\
M_{I}=& 0.4711-1.3118\log{P}+0.2053\log{Z},\nonumber
\end{align}    
where $\log{Z}$ is given by \citep{sala93,cate04} 
\begin{align*}
\log{Z}=& [Fe/H]+\log\left(0.638f+0.362\right)-1.765. 
\end{align*}
\begin{figure}
\begin{center}
\includegraphics[width=0.5\textwidth,keepaspectratio]{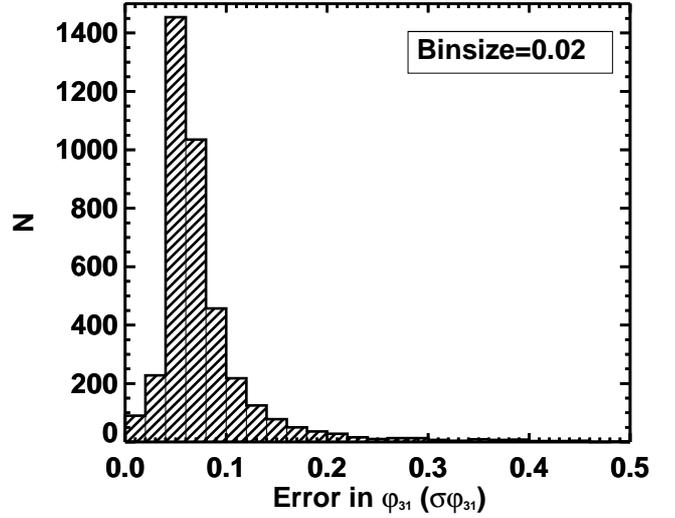}
\caption{Histogram plot of the distribution of standard errors in $\sigma_{\phi_{31}}$ for the $I$-band data.}
\label{ephi31}
\end{center}
\end{figure}
The $M_{V}$ relation was taken from \citet{cate08} and that of $M_{I}$ 
from \citet{cate04}. $f=10^{[\alpha/Fe]}$ denotes the enhancement 
$\alpha$-elements with respect to iron. For the SMC 
$\left<[\alpha/Fe]\right>=0.2$ \citep{varg14}. Therefore, 
$f=10^{0.2}=1.585$. Here $[Fe/H]$ is the metallicity in the 
\citet[hereafter ZW84]{zinn84} scale. The value of $[Fe/H]$ can be calculated 
using the most accurate and up-to-date empirical $[Fe/H]-P-\phi_{31}$ 
non-linear relation of \citet{neme13}. This relation makes use of the highly 
accurate and precise Kepler $Kp$-band $26$ RRab data with good photometric 
light curves along with metallicities obtained from high resolution 
spectroscopic measurements:
\begin{eqnarray}
[Fe/H]_{N13} = (-8.65 \pm 4.64) + (-40.12 \pm 5.18)P+~~~~\nonumber \\
(5.96\pm1.72)\phi_{31}^{kp}+ (6.27 \pm 0.96)\phi_{31}^{kp} \times P+~~~\nonumber \\  
(-0.72 \pm 0.17)(\phi_{31}^{kp})^2,~~~\sigma = 0.084
\label{eq:neme13}
\end{eqnarray}
It should be noted that the value of $\phi_{31}^{kp}$ used in the above 
relation is obtained from the Fourier sine decomposition \citep{neme13} and is 
in the Kepler photometric $Kp$-band. The above empirical relation may not 
provide accurate value of $[Fe/H]_{N13}$ for an individual RRab star but is 
suitable for statistical analysis of a large number of RRab stars 
\citep{skow16}. From the very extensive study by \citet{skow16}, it has been 
demonstrated that the $[Fe/H]_{\text{N13}}$ values obtained using the above 
relation with the metallicity-dependent and metallicity-independent 
transformations involving $(\phi_{31}^{I} \rightarrow \phi_{31}^{V})$ are 
different. The differences for individual RRab stars may  reach up to $0.3$ 
dex and are typically lower than $0.1$ dex. But in a statistical study 
involving a large sample of RRab stars these differences get reduced to 
smaller values. The relation given by equation~(\ref{eq:neme13}) was based on 
the metallicity scale of \citet[hereafter JK95]{jurc95}. The above relation in 
$K_{p}$ can be applied to the $V$-band data using the following inter-relation 
\citep{jeon14,skow16}
\begin{equation}  
\phi_{31}^{V}=\phi_{31}^{Kp}-(0.174\pm0.085).
\label{eq:jeon14}
\end{equation} 
}
Since the $I$-band light curve of OGLE-IV RRab stars contains more number of 
data points with better phase coverage, $\phi_{31}$ determined from the 
$I$-band will be more precise and accurate as compared to that of the 
$V$-band. We use these $\phi_{31}^{I}$ values to convert them into their 
corresponding $V$-band values using the more accurate metallicity-independent 
inter-relation obtained by \citet{skow16}:    
\begin{eqnarray}
\phi_{31}^{V}= (0.122\pm0.017)(\phi_{31}^{I})^2-(0.750\pm 0.187)\phi_{31}^{I}+~~~~~~~~~\nonumber \\ 
(5.331\pm0.523),~~~~~~~~~\sigma_{\rm fit}=0.004.
\label{eq:skow16}
\end{eqnarray}
Using the inter-relations from equations~(\ref{eq:jeon14}) and (\ref{eq:skow16})
we obtain the metallcities of the individual RRab stars from 
equation~(\ref{eq:neme13}). The metallicity values obtained from 
equation~(\ref{eq:neme13}) are in \citet{jurc96} scale, which can be 
transformed into the metallicity scale of \citet{zinn84} using the following 
relation from \citet{jurc95}:
\begin{equation}         
\label{eq:zw84}
[Fe/H]=\frac{[Fe/H]_{N13}-0.88}{1.431}.
\end{equation}
Although there are other $[Fe/H]-P-\phi_{31}$ relations applicable for the 
$V$- and $I$-band data, respectively developed by \citet{jurc96} and 
\citet{smol05}, the $[Fe/H]$ values obtained using these relations are found to 
systematically overestimate as well as underestimate their values towards the 
low and high metallicity ends. One of the important advantages of using the 
\citet{neme13} relation is that it corrects these problems in metallicity 
calculations \citep{skow16}.

Nonetheless it has been observed from the studies of \citet{deb15} and 
\citet{skow16} that the formal errors on $[Fe/H]$ obtained by applying the 
propagation of error formula on the \citet{neme13} metallicity relation 
given by equation~(\ref{eq:neme13}) are exceedingly large. However the 
consequences of this effect are not important in a statistical analysis of a 
large sample of stars done in the present study. Following \citet{skow16} the 
errors in the [Fe/H] values of individual RRab stars using 
equation~(\ref{eq:neme13}) are obtained by carrying out Monte Carlo simulations 
assuming that the distributions of the coefficients in the 
equation~(\ref{eq:neme13}) are Gaussian. A small random Gaussian noise of $0.01$ 
is  added to each of the four coefficients in the equation~(\ref{eq:neme13}). 
Monte Carlo simulations are performed with $1000$ iterations each time 
calculating the $[Fe/H]_{N13}$ values from the randomly generated 
coefficients.  This iterative procedure applied to obtain each of the 
$[Fe/H]_{N13}$ values helps us to build up the $[Fe/H]_{N13}$ distribution. 
Gaussian distribution with parameters $\mu$ and $\sigma$ is then fitted to 
the histogram which yield the corresponding true value of $[Fe/H]_{N13}$ and 
its associated error for an individual RRab star. 
\section{Application of the technique to the OGLE-IV database}
\label{selection}
{\bf 
To determine the values of reddening-free distance modulus $(\mu_{0})$ and 
reddening ($E(B-V)$) of an individual RRab star using the technique as 
described in Section~\ref{methodology}, we apply the following steps in order:
\begin{enumerate}
\item  Conversion of $\phi_{31}^{I}$ into $\phi_{31}^{V}$ using equation~(\ref{eq:skow16})
\item $\phi_{31}^{V}$ is then converted into $\phi_{31}^{Kp}$ using 
equation~(\ref{eq:jeon14})  
\item $[Fe/H]_{N13}$ and its error $\sigma_{[Fe/H]_{N13}}$ are obtained using 
equation~(\ref{eq:neme13}) with Monte Carlo simulations as described above
\item $[Fe/H]_{N13}$ is then converted into the ZW84 metallicity scale 
$[Fe/H]$ using equation~(\ref{eq:zw84})
\item Absolute magnitudes $M_{V}$ and $M_{I}$ are determined using 
equation~(\ref{eq:absmag})
\item $\mu_{V}$ and $\mu_{I}$ are obtained using equation~(\ref{eq:mu})
\item $E(B-V)$ and $\mu_{0}$ are determined from equation~(\ref{simul})     
\item $D$ is calculated using equation~(\ref{eq:distance}).
\end{enumerate} 
It has already been mentioned that the values of $\phi_{31}$ as given in the 
OGLE-IV database are for the $I$-band and are obtained from the Fourier cosine 
decomposition. On the other hand, we have seen that the \citet{neme13} 
relation is derived based on the value of $\phi_{31}$ in the Kepler 
photometric $Kp$-band and is obtained from a Fourier sine decomposition. 
This fact also has to be kept in mind while trying to obtain metallicity 
values using the \citet{neme13} relation with the values of $\phi_{31}$ as 
given in the database. Therefore if the values of $\phi_{31}$ as given in the 
database were used we would be required to convert them first into their 
corresponding values as those obtained from a Fourier sine decomposition by 
adding a value of $\pi$ radians to them and finally convert these values to the 
$Kp$-band. Since we do not use the values of $\phi_{31}$ as given in the 
database, addition of $\pi$ is not required in our case. Let us demonstrate 
the above steps for the case of star with OGLE ID: OGLE-SMC-RRLYR-0001. As 
given in the database, $P=0.5588145$d. From the Fourier sine decomposition 
technique in the present paper, we obtain $\overline{m}_{V}=19.584 \pm 0.005$ 
mag, $\overline{m}_{I}=19.067 \pm 0.002$ mag, $\phi_{31}^{I}=5.360 \pm 0.060$. 
The above steps applied to this star yield
\begin{enumerate}
\item $\phi_{31}^{V}=4.816$ rad 
\item $\phi_{31}^{Kp}=4.990$ rad
\item $[Fe/H]_{N13}=-1.773\pm0.266$ dex
\item $[Fe/H] = -1.855\pm0.186$ dex
\item $M_{V} = 0.558\pm0.043 $ mag, $M_{I}=0.088\pm 0.038$ mag
\item $\mu_{V}=19.026\pm 0.043 $ mag and $\mu_{I}=18.979\pm 0.038$ mag  
\item $E(B-V)=0.038\pm 0.046$ mag and $\mu_{0}=18.905\pm 0.098$ mag
\item $D=60.397\pm 2.726$~kpc. 
\end{enumerate}   
Now we calculate all the aforementioned parameters using the values of the 
mean magnitudes ($\overline{m}_{V},\overline{m}_{V}$) and Fourier phase 
parameter $\phi_{31}^{I}({\rm cosine})$ as given in the OGLE-IV database.  
The values of the parameters are: $P=0.5588145$d, $\overline{m}_{V}=19.564$ 
mag, $\overline{m}_{I}=19.050$ mag, $\phi_{31}^{I}({\rm cosine})=2.199$ 
(obtained from the Fourier cosine decomposition). To convert 
$\phi_{31}^{I}({\rm cosine})$ to the corresponding value to be obtained 
from the Fourier sine decomposition, we have to add $\pi$ to it. Therefore, we 
have $\phi_{31}^{I} \rightarrow \phi_{31}^{I}({\rm cosine}) +\pi = 2.199+\pi= 5.341$ with the condition that 
$\phi_{31}^{I} \in [0,2\pi]$. The above steps applied to this star yield     
\begin{enumerate}
\item $\phi_{31}^{V}=4.805$ rad 
\item $\phi_{31}^{Kp}=4.979$ rad
\item $[Fe/H]_{N13}=-1.798$ dex
\item $[Fe/H] = -1.872$ dex
\item $M_{V} = 0.554$ mag, $M_{I}=0.084$ mag
\item $\mu_{V}=19.010 $ mag and $\mu_{I}=18.966$ mag  
\item $E(B-V)=0.036$ mag and $\mu_{0}=18.896$ mag
\item $D=60.135$~kpc. 
\end{enumerate}   
We have thus demonstrated that the $E(B-V)$ and $D$ values of an RRab star 
determined using the light curve parameters obtained from the Fourier sine 
decomposition technique in the present paper are consistent with their 
corresponding values obtained using the parameter values as given in the 
OGLE-IV database which have been obtained from the Fourier cosine 
decomposition. Having demonstrated the application of the technique to a single
RRab star, the $E(B-V)$ and $D$ values of  all the $3522$ sample of RRab 
stars are determined in a similar way using the Fourier sine decomposition 
method in the present study. }            
\section{Comparison of Distance and Reddening obtained using other Relations}
\label{comp_other}
\begin{figure*}
\vspace{0.02\linewidth}
\begin{tabular}{cc}
\vspace{+0.01\linewidth}
  \resizebox{0.50\linewidth}{!}{\includegraphics*{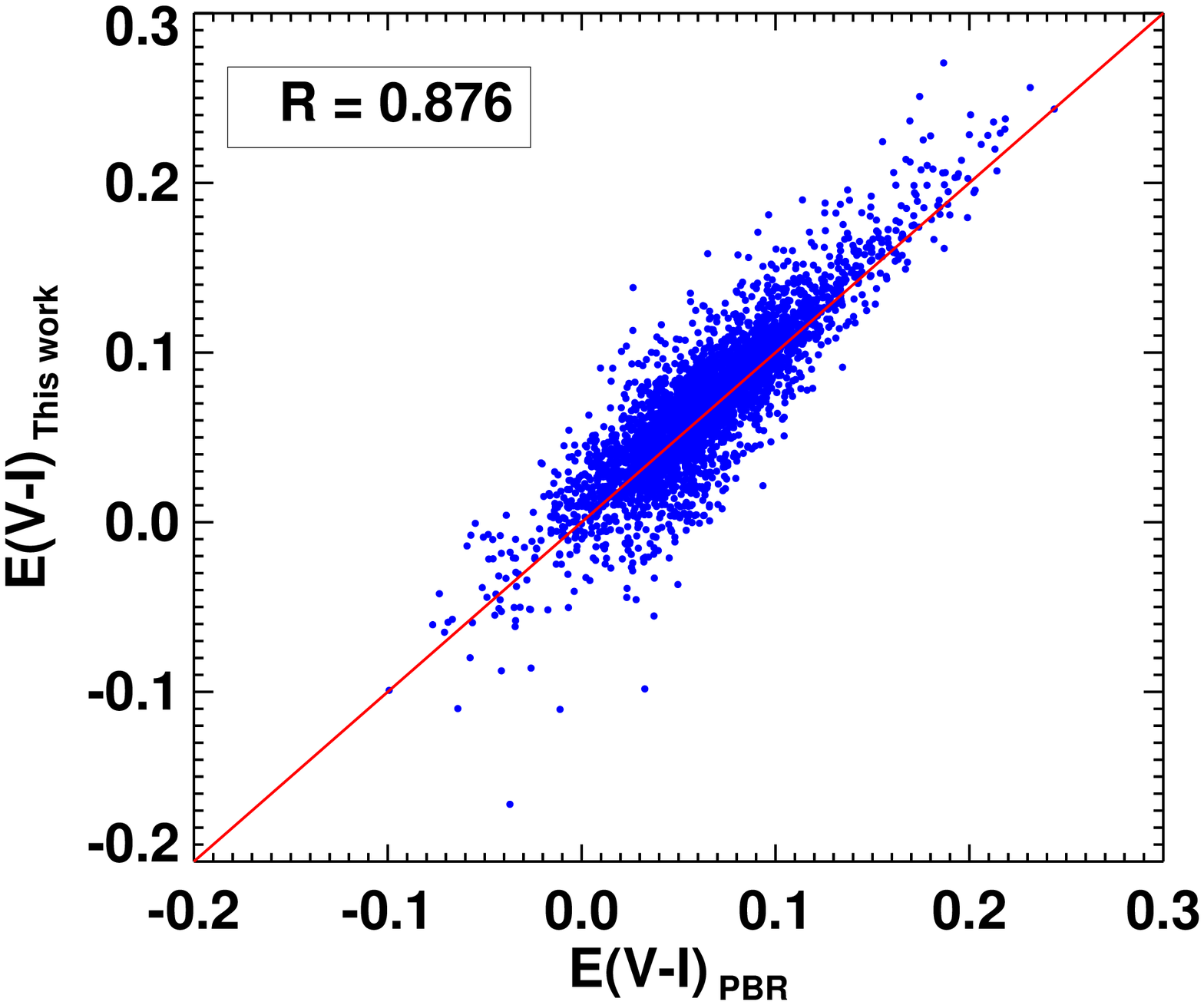}}&
  \resizebox{0.50\linewidth}{!}{\includegraphics*{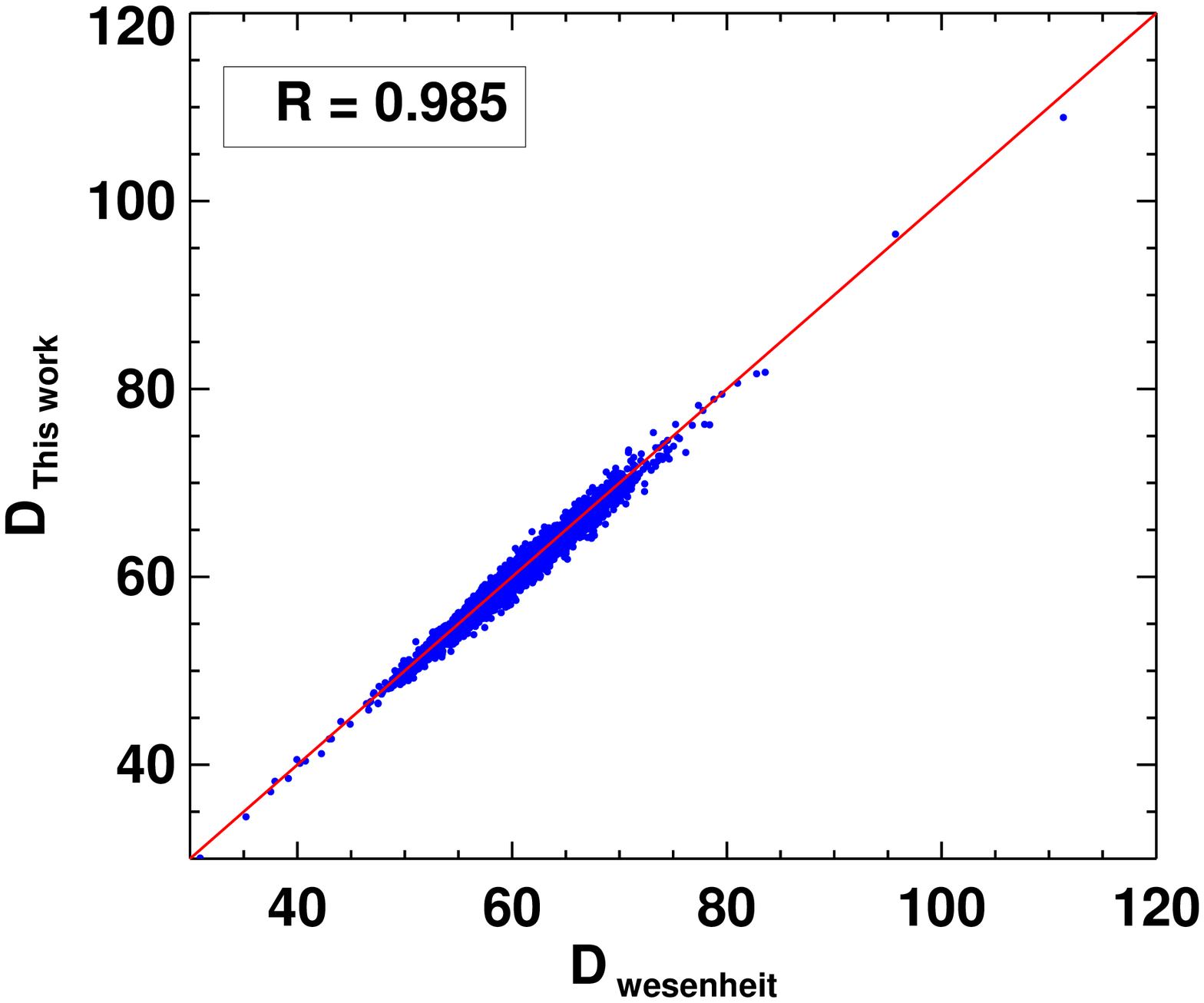}}\\
\vspace{-0.04\linewidth}
\end{tabular}
\caption{{\bf Reddening values and distances obtained using the methodology  as 
described in Section~(\ref{methodology}) $vs$ their corresponding values 
calculated using equations~(\ref{color_pbr}) and (\ref{dist_w}) are plotted in 
$1:1$ scatter plots. $R$ denotes the correlation coefficient.}}
\label{comp_dredd}
\end{figure*}

{\bf
In this Section we compare the values of reddening and distance obtained for
the $3522$ RRab stars with their corresponding values obtained using other
empirical and theoretical relations available in the literature. There exists
an empirical relation which connects the intrinsic colour $(V-I)_{0}$ of an 
RRab star with its $V$-band amplitude $A_{V}$ and period $P$  \citep{pier02}:
\begin{align}
(V-I)_{0}=& (0.65\pm0.02)-(0.07\pm0.01)A_{V}+(0.36\pm0.06)\log{P},
\end{align}
where $\sigma_{\text{fit}}=0.02$.
The $V$-band amplitudes ($A_{V}$) are obtained from the following relation 
\citep{deb10}:
\begin{align}
A_{V}=(1.500\pm0.040)A_{I}+(0.071\pm 0.019).
\end{align}     
Reddening $E(V-I)$ is defined as 
\begin{align}
\label{color_pbr}
E(V-I)=(\overline{m}_{V}-\overline{m}_{I})-(V-I)_{0}.
\end{align}
Here $A_{I}$, $\overline{m}_{V}$ and $\overline{m}_{I}$ are determined from 
the Fourier sine decomposition method as described in Section~\ref{data}.    
Distance of an individual RRab star is obtained from the Wesenheit distance 
modulus given by \citep{jacy17} 
\begin{align*}  
\mu_{W}=W_{\rm obs}-W_{\rm abs},
\end{align*}
where $W_{\rm obs}$ and $W_{\rm abs}$ denote the observed and theoretical 
absolute  Wesenheit indices, respectively, given by \citep{skow16,jacy17}
\begin{align}
W_{\rm obs}=\overline{m}_{I}-1.55(\overline{m}_{V}-\overline{m}_{I})
\end{align}    
and
\begin{align}
\label{eq:brag15}
W_{\rm abs}=a_{W}+b_{W}\log{P}+c_{W}\left([Fe/H]_{C}+0.04\right).
\end{align}  
Here $a_{W}=-1.039\pm 0.007$, $b_{W}=-2.524\pm0.021$ and 
$c_{W}=0.147\pm 0.004$ \citep{brag15}. $[Fe/H]_{C}$ denotes the metallicity 
value of an RRab star in the \citet{carr09} scale given as follows 
\citep{kapa11}
\begin{align*}
[Fe/H]_{C}=& \left(1.001\pm 0.050\right)[Fe/H]_{N13}-(0.112\pm0.077). 
\end{align*}  
The relation~(\ref{eq:brag15}) is the theoretical Period-Wesenheit-Metallicity
(PWZ) relation obtained by \citet{brag15} and has also been used by 
\citet{jacy17} to calculate the distances of SMC RRab stars in the OGLE-IV 
database. One of the important advantages of using Wesenheit index in the 
distance determination is that by virtue of its construction it is from the 
interstellar extinction, assuming that the reddening law is known 
\citep{mado82}. Distances are then calculated using:
\begin{align}
\label{dist_w}
D [\rm pc]=& 10^{\frac{\left(\mu_{W}+5\right)}{5}}.
\end{align}
Reddening values and distances obtained using equations~(\ref{color_pbr}) \& 
(\ref{dist_w}) are denoted by $E(V-I)_{\rm PBR}$ and $D_{\rm Wesenheit}$, 
respectively.  Reddening values and distances obtained in the present study 
using the methodology as described in Section~(\ref{methodology}) against 
their corresponding values calculated using equations~(\ref{color_pbr}) and 
(\ref{dist_w}) are plotted in $1:1$ scatter plots as shown in 
Fig.~(\ref{comp_dredd}). Although
the relations given by equations~(\ref{color_pbr}) and (\ref{dist_w})  are 
obtained from entirely different calibrations, their consistency with the 
present study demonstrate an independent and robust proof of validity of these
relations.  Mean values of the reddening and distance to the SMC using 
equations~(\ref{color_pbr}) and (\ref{dist_w}) are found to be 
$0.077\pm 0.040$ mag and $61.092\pm4.145$ kpc, respectively. These values are 
quite consistent with their corresponding mean values of $0.066\pm 0.036$ mag
and $60.735\pm4.143$ kpc, respectively, within the quoted uncertainties as 
discussed in the next Section. }          
\section{Mean Distance and Reddening to the SMC}
\label{distance}
\begin{figure*}
\vspace{0.02\linewidth}
\begin{tabular}{cc}
\vspace{+0.01\linewidth}
  \resizebox{0.50\linewidth}{!}{\includegraphics*{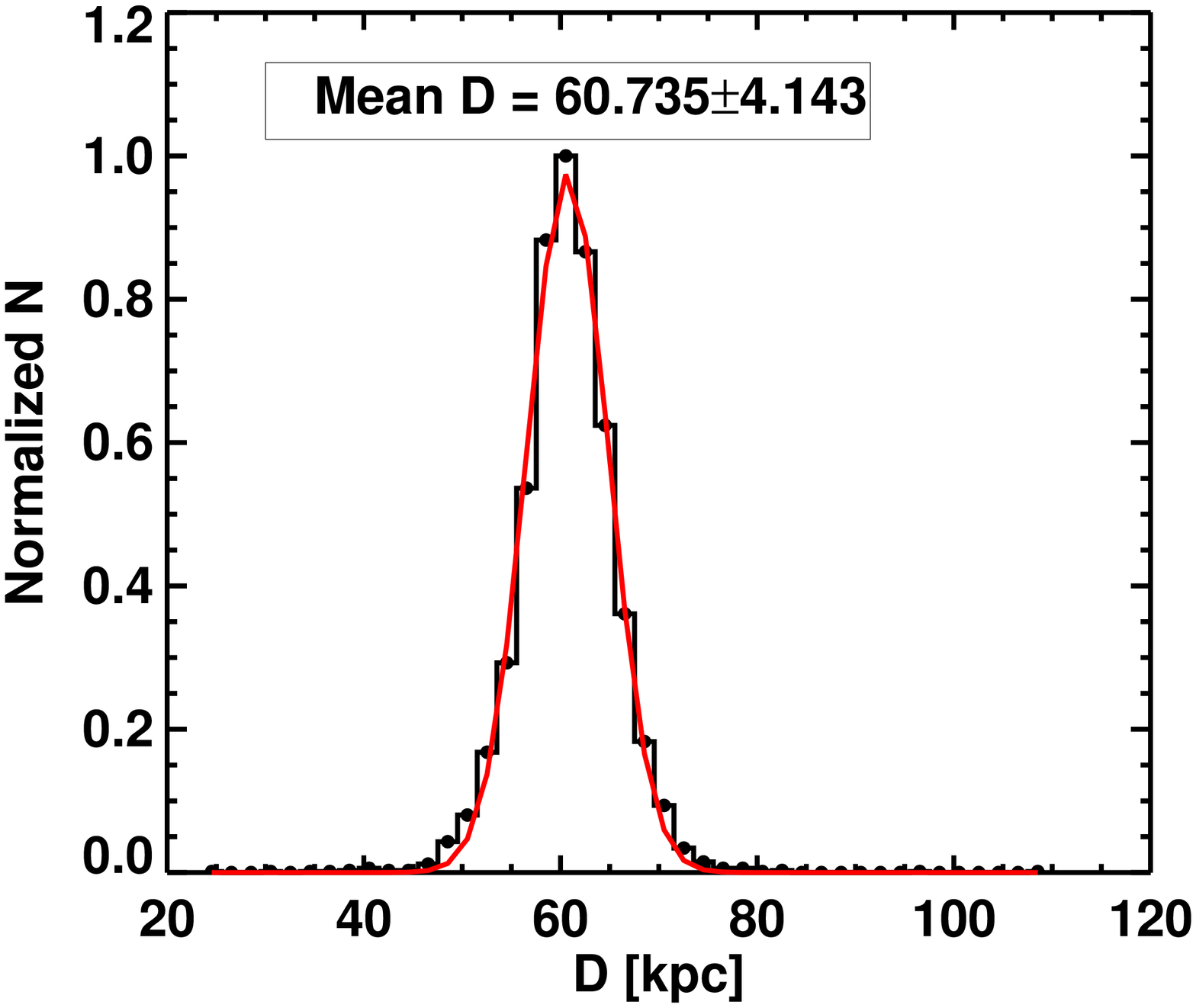}}&
  \resizebox{0.50\linewidth}{!}{\includegraphics*{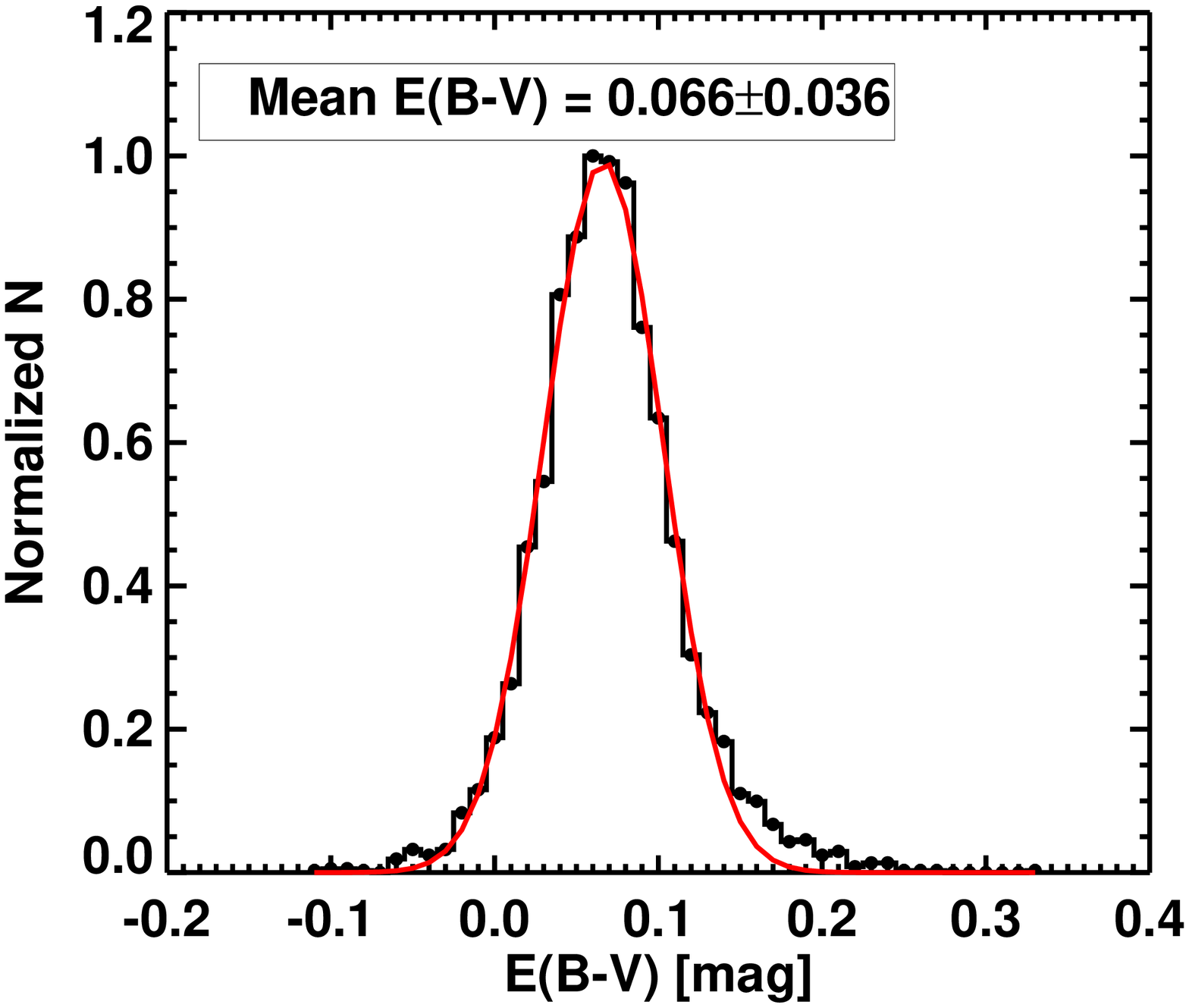}}\\
\vspace{-0.04\linewidth}
\end{tabular}
\caption{{\bf Normalized distance and reddening distributions of the selected 
sample of OGLE-IV SMC $3522$ RRab stars. Red colour solid lines in both the 
two cases represent the three parameter Gaussian fit of their respective 
distributions.}}
\label{normalized}
\end{figure*}
\begin{figure*}
\vspace{0.02\linewidth}
\begin{tabular}{ccc}
\vspace{+0.01\linewidth}
  \resizebox{0.32\linewidth}{!}{\includegraphics*{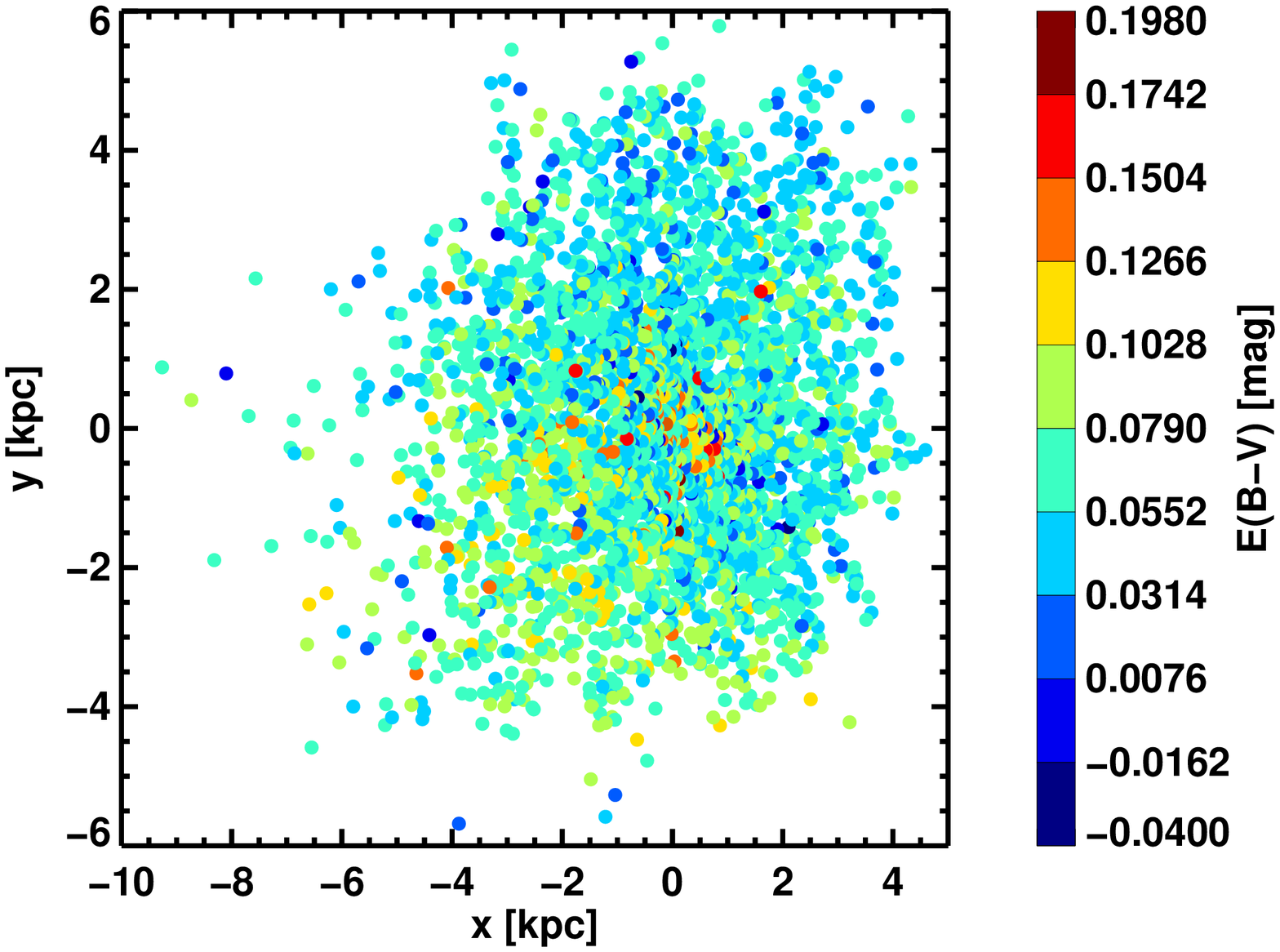}}&
    \resizebox{0.32\linewidth}{!}{\includegraphics*{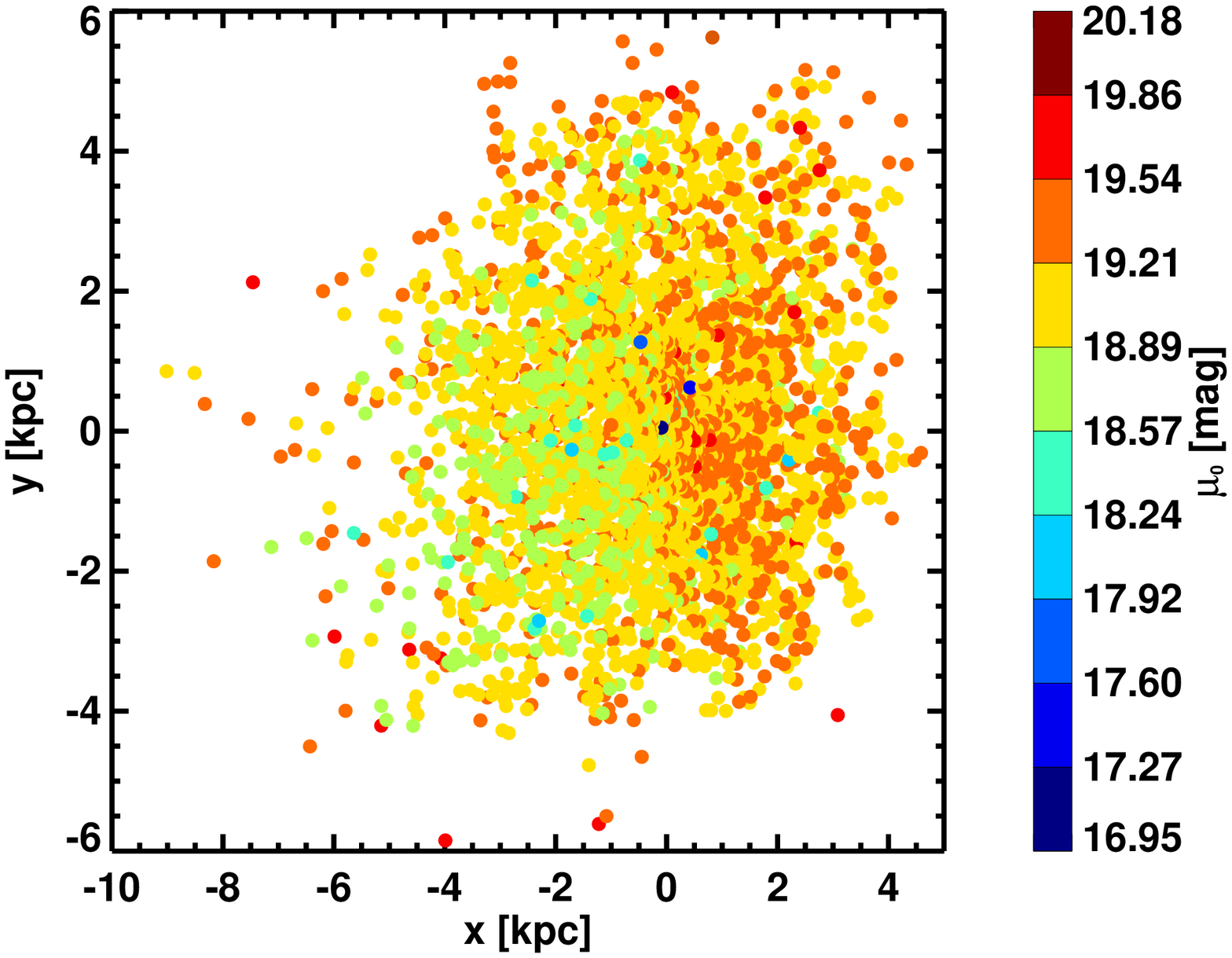}} &
  \resizebox{0.32\linewidth}{!}{\includegraphics*{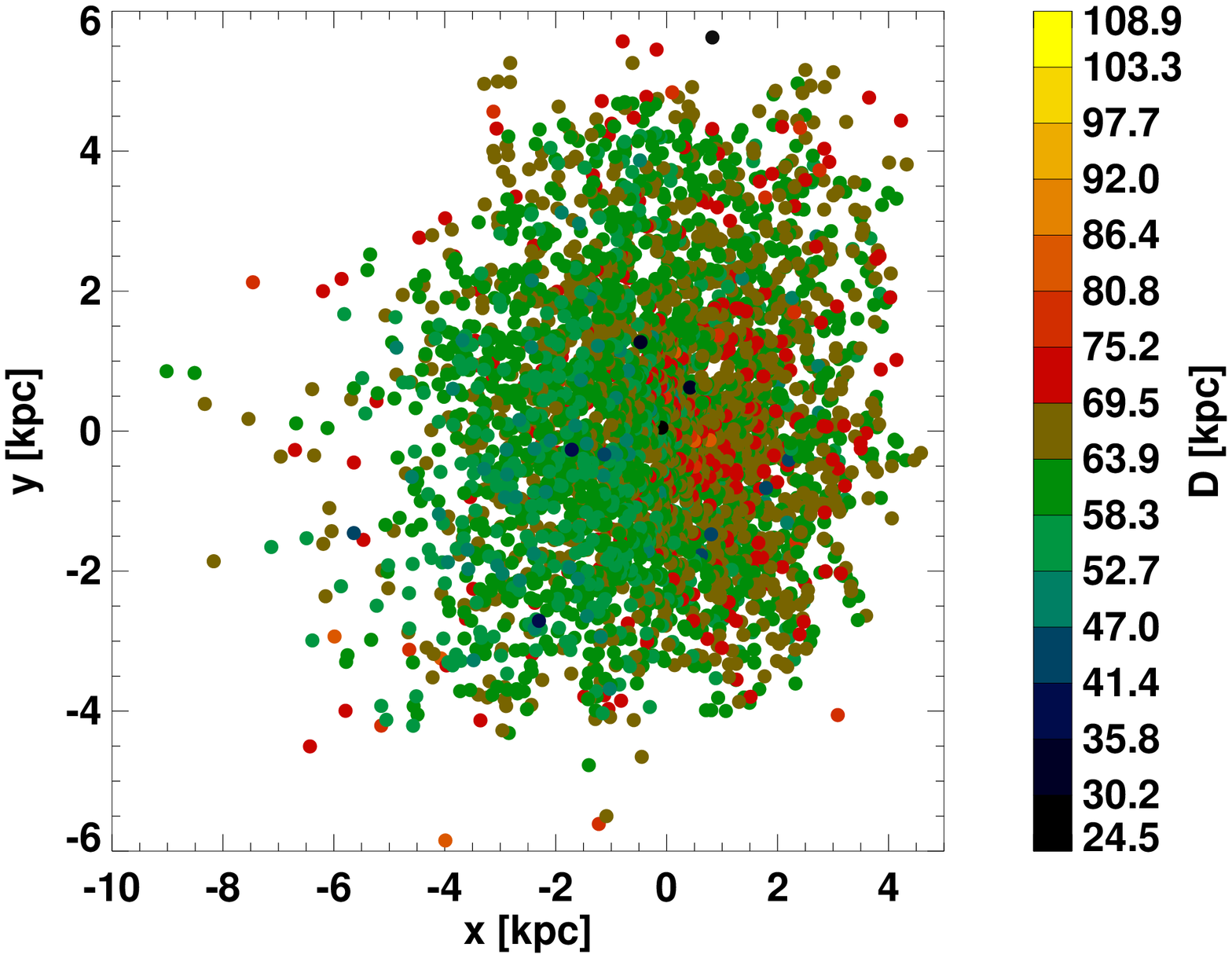}}\\
\vspace{-0.04\linewidth}
\end{tabular}
\caption{{\bf Two-dimensional colour bar plots of the reddening values ($E(B-V)$), 
true distance moduli $(\mu_{0})$ and true distances $(D)$ of each of the 
$3522$ RRab stars. Here $(x,y)$ represent the Cartesian coordinates.}}
\label{cbar}
\end{figure*}
\begin{figure}
\begin{center}
\includegraphics[width=0.5\textwidth,keepaspectratio]{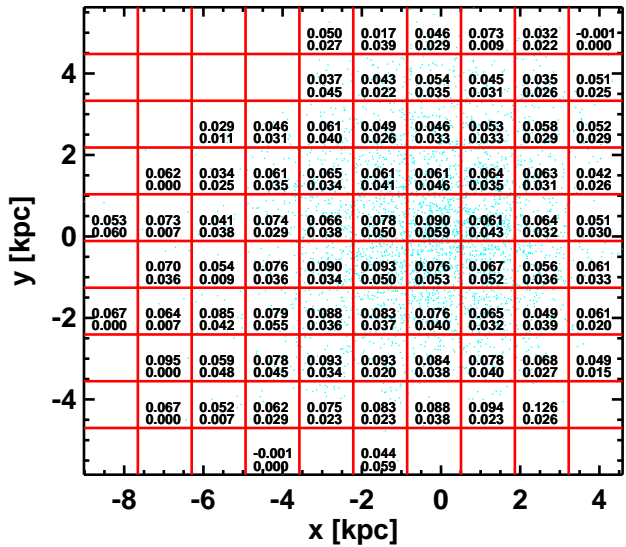}
\caption{{\bf Reddening distribution $E(B-V)$ of the {\bf $3522$} RRab stars in the SMC. 
$E(B-V)$ values are binned on a $10\times 10$ coordinate grid. In each bin, 
average reddening values and their associated errors calculated are shown in 
each grid box.}}
\label{reddening_dens}
\end{center}
\end{figure}
\begin{figure}
\begin{center}
\includegraphics[width=0.5\textwidth,keepaspectratio]{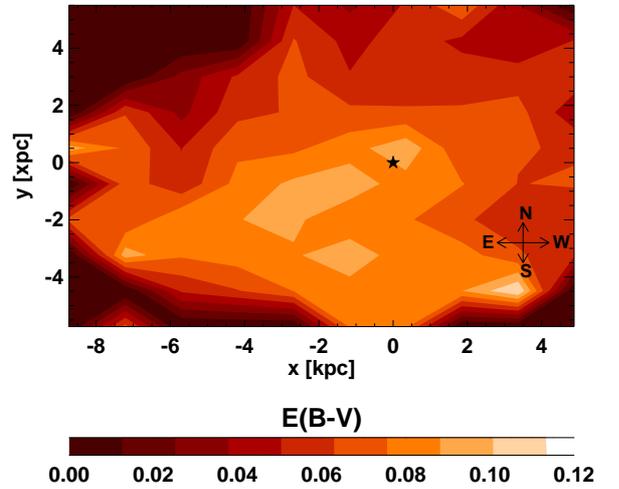}
\caption{{\bf The reddening map of the SMC derived from the SMC RRab stars. The map
is produced by computing the average reddening on a $10\times10$ grid in 
$(x,y)$ coordinates. The location of the centre of the SMC is shown as a star 
symbol.}}
\label{map}
\end{center}
\end{figure}

{\bf
The distance and reddening distributions for the selected sample of $3522$ 
RRab stars are shown in Fig.~\ref{normalized}. The distribution functions 
approximate nearly to those of Gaussian profiles. We have fitted a 
three-parameter Gaussian function to the distance and reddening distribution 
of these RRab stars. The following values are obtained $D=60.735\pm 4.143$ kpc 
and $E(B-V)=0.066\pm 0.036$ mag. 
Here the uncertainties represent the spread in the population rather than the 
standard deviation of the mean. On the other hand, weighted averages of the 
distance and reddening of each of the RRab stars yield the mean distance and 
reddening as $59.845\pm0.046$ kpc and $0.071 \pm 0.001$ mag, 
respectively, where the uncertainties represent the standard errors of the 
mean values. From reddening distribution plot, it can be seen that for some of 
the stars, we get unphysical negative reddening values. The number of stars 
with $E(B-V) <0$ are found to be $149$ which is a very small number 
($\sim 4\%$) as compared to the total number of stars in the present study. 
The mean value of reddening of these $149$ stars is found to be 
$-0.023\pm0.022$ mag which is statistically not significant and is consistent 
with zero within the uncertainties. If the stars with negative 
$E(B-V)$ are set to zero, we get the mean vaues the distance and reddening of 
the SMC as $D=60.723\pm4.080$ kpc and $E(B-V)=0.064\pm0.039$ mag, 
respectively. One of the causes of getting the negative reddening values for 
these $149$ stars may be due to the unreliable estimates of their $V$-band 
mean magnitudes which are underestimated due to the noisy and poorly sampled 
data points of their light curves or may be due to the propagated 
uncertainties in their 
calculated values. These are the two specific cases in which the reddening 
estimations using equation~(\ref{simul}) are likely to yield negative values. 
Stars  with negative $E(B-V)$ values have the values of $[Fe/H]$  in the range 
$-2.68 \le [Fe/H] \le -1.26$ dex. Mean values of the periods and metallicities 
of these $149$ stars are found to 
$P_{E(B-V)<0}=0.650540\pm 0.0689168~\text{d}$ and 
$[Fe/H]_{E(B-V) <0 } = -1.935\pm0.290~\text{dex}$. The values of the mean 
period and metallicity of the stars with negative reddening values are 
respectively higher and lower, as compared to the overall mean values of the 
total sample $P = 0.595629\pm 0.0513323~{\rm d}$ and 
$[Fe/H] = -1.859\pm 0.272~\text{dex}$. We have also studied 
whether there is a trend of these negative reddening values as a function of 
their periods as well as metallicities but we do not find any clear trend 
or correlation. In the analysis, we do not ignore those stars with negative 
reddening values keeping in view of their uncertainties else otherwise this 
will skew the distribution towards positive reddening values and may lead to 
biases \citep{mura17}.   

From Fig.~\ref{devi}, it should be noted that the distance and reddening 
values obtained for the final sample of $3522$ RRab stars are anticorrelated.
The expected behaviour is totally in contrast to what is observed: stars 
located at larger distances should have higher values of reddening 
\citep{niko04}. This reflects an independent and unbiased determination of 
these two quantities using the methodologies described in 
Section~\ref{methodology}. Fig.~\ref{cbar} depicts the two-dimensional colour 
bar plots of the distribution of the reddening values $E(B-V)$, true distance 
modulus $\mu_{0}$ and true distances $D$ for each of the $3522$ RRab stars. 
Reddening distribution of the SMC RRab stars is shown in 
Fig~\ref{reddening_dens}. $E(B-V)$ values are binned on a $10\times10$ 
coordinate grid. The average $E(B-V)$ values and their associated errors are 
given in each bin. The reddening map of the SMC derived from the SMC RRab 
stars is shown in Fig.~\ref{map}. The map is produced by computing the average 
reddening on a $10\times10$ grid in $(x,y)$ coordinates. 
\begin{figure}
\begin{center}
\includegraphics[width=0.5\textwidth,keepaspectratio]{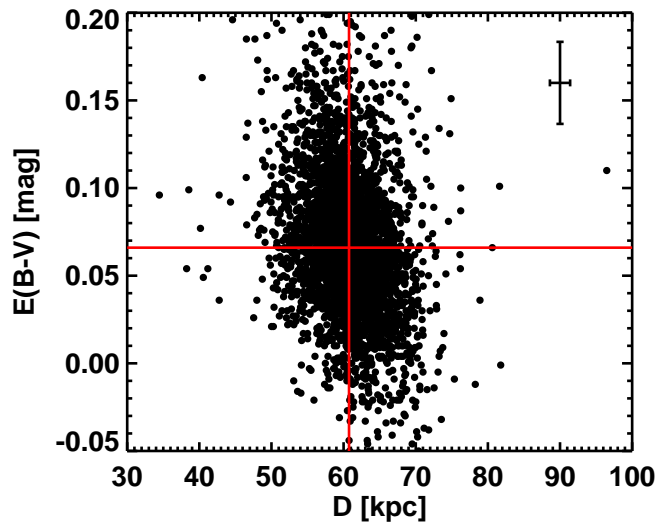}
\caption{{\bf Reddening values vs. distance for the final sample of $3522$ RRab 
stars. Note that these two values are anticorrelated. Typical size of the 
error bars for individual points are shown in the upper right corner of the 
plot. The intersection of the two solid lines (red) represent the central values
of these parameters of the SMC.}}
\label{devi}
\end{center}
\end{figure}
Four prominent zones of high internal reddening can be located from the map:
one to the south-western part, another one to the north and two others to the 
south-eastern parts with respect to the centre of the SMC. From the reddening 
map, one can see that the regions located to the south-eastern part of the 
centre of the SMC have the highest average values of reddening as compared to 
the other parts. It can also be seen from the map that the southern part of 
the SMC has relatively high internal reddening zones as compared to its 
northern part. One of the highest internal reddening zones is seen to be 
located near to south-western edge of the SMC.

In a study done using the red clump (RC) stars 
from the OGLE-II database, \citet{subr09} found that the western 
side of the SMC has high internal reddening. On the other hand using the RC 
stars and RRab stars of the SMC from OGLE-III database, \citet{hasc11} 
also found that the south-western part of the SMC has higher reddening values.
The reddening maps by \citet{hasc11,subr12} and the recently obtained reddening 
map by \citet{mura17} are in good agreement with our findings.  The mean 
reddening of the SMC $E(B-V)=0.066\pm 0.036$ mag 
determined using the present sample of $3522$ OGLE-IV SMC RRab stars is found 
to be in good agreement with the values of $E(B-V)=0.048\pm 0.039$ mag, 
$E(B-V)=0.054\pm0.029$ mag and $E(V-I)=0.07\pm 0.06$ mag, respectively obtained 
from the \citet{zari02} reddening map using the OGLE-III SMC RRab stars by 
\citet{deb15}, from the analysis of $48$ Cepheids by \citet{cald86} and from 
the analysis $1529$ OGLE-III SMC RRab stars by \citet{hasc11}. 
The conversion relation between $E(V-I)$ and $E(B-V)$ is given by 
$E(V-I)=1.26E(B-V)$. On the other hand, the mean distance to the 
SMC $D=60.735\pm 4.143$ kpc obtained in the present study is quite consistent 
with the recent estimates of distance determination to the SMC: 
$62.1\pm 1.9$ kpc by \citet{grac14}, $61.09\pm1.47$ by \citet{inno13} using 
the $2571$ fundamental mode Cepheids (FU) in the largest NIR $JHK$ band 
datasets. Furthermore using the NIR $JHK$ datasets for the first overtone Cepheids (FO), \citet{inno13} found a true distance modulus of the SMC to be 
$19.12\pm 0.13$ mag. This corresponds to a true distance to the SMC of 
$66.68\pm 1.73$ kpc which is an overestimation of the distance to the 
SMC when compared with the other values in the literature. \citet{inno13} 
cited this overestimation of the SMC distance due to the lack of precise 
trigonometric parallax determinations of the Galactic FO Cepheids for 
distance calibration.   Although we have compared the SMC mean distance and 
reddening value obtained from the Gaussian fit with their respective 
values available in the literature, these values have been refined to get 
their intrinsic mean and intrinsic spread using a robust maximum likelihood 
estimation method as discussed in the next section. }            
\section{Error Estimation}
\label{error}
\begin{figure}
\begin{center}
\includegraphics[width=0.5\textwidth,keepaspectratio]{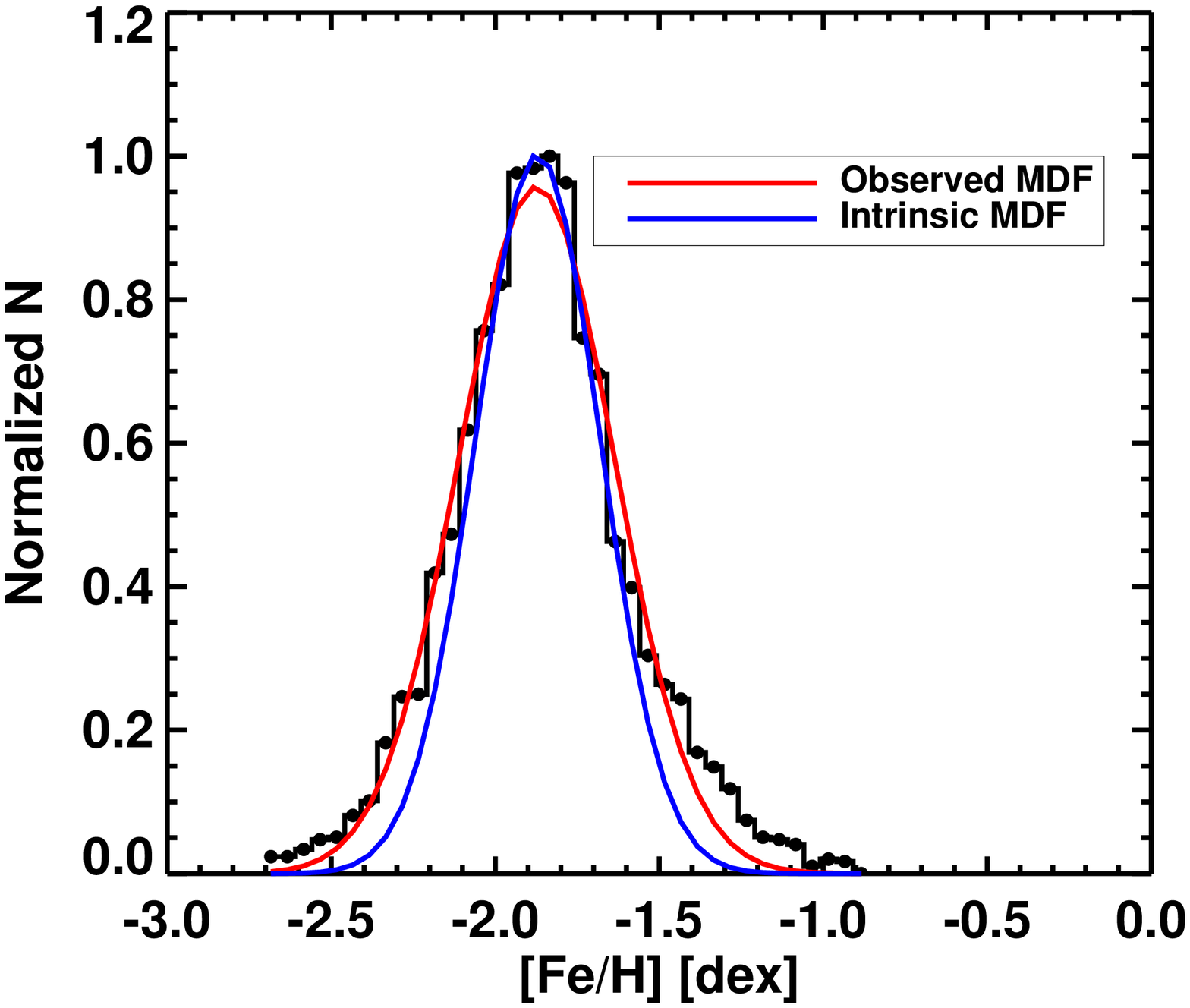}
\caption{{\bf The red colour solid line shows the normalized metallicity 
distribution function (MDF) with the individual metallicity values convolved 
with the individual Gaussian uncertainties while the blue solid line depicts 
the intrinsic Gaussian distribution.}}
\label{mdf}
\end{center}
\end{figure}

Each of the parameters $\sigma_{\overline{m}_{V,I}}$ and $\sigma_{\phi_{31}}$ 
obtained from the light curve analysis of RRab stars using the Fourier 
decomposition technique will contain standard errors. This will result into 
the  errors in the determination of other parameters dependent on them 
according to the propagation of error formula \citep{bevi03}. Apart from that, 
each of the observed parameters determined using the empirical and theoretical 
calibrations will contain systematic errors. Using the propagation of errors 
formula when the number of observations are large, one can approximately find 
that the error in the measurement of $M_{V}$ is $\sigma_{M_{V}}=0.23\sigma_{[Fe/H]}$, $M_{I}$ is $\sigma_{M_{I}}=0.2503\sigma_{[Fe/H]}$, $\mu_{I}$ is $\sigma_{\mu_{I}}=\sqrt{\sigma_{\overline{m}_{I}}^{2}+\sigma_{M_{I}}^{2}}$ ,  $\mu_{V}$ is $\sigma_{\mu_{V}}=\sqrt{\sigma_{\overline{m}_{V}}^{2}+\sigma_{M_{V}}^{2}}$, $E(B-V)$ is $\frac{\sqrt{\sigma_{\mu_{V}^2}+\sigma_{\mu_{I}^2}}}{R_{V}-R_{I}}$, $\mu_{0}$ is $\sigma_{\mu_{0}}=\sqrt{\sigma_{\mu_{I}}^2+R_{I}^{2}\sigma_{E(B-V)}^{2}}$, $D$ is $\frac{D\sigma_{\mu_{0}}}{2.17147}$. The total uncertainty in 
$[Fe/H]$ values are the errors obtained from the Monte Carlo simulations 
as well as the quoted systematic uncertainty of $0.084~\text{dex}$ as 
mentioned in equation~(\ref{eq:neme13}). These uncertainties are added 
quadratically resulting into a total mean uncertainty of 
$\sigma_{[Fe/H]}=0.21~\text{dex}$. In order to find a better estimate of the 
mean and intrinsic spread in the metallicity distribution we use the maximum 
likelihood estimation method. One of the advantages of maximum likelihood 
method is that it treats each of the observations independently thus allowing 
the parameter estimations free from any cumulative systematic errors. The 
observed metallicity distribution function (MDF) approximates that of the 
nearly Gaussian profile.  The resulting MDF can be thought of as the 
convolution of an intrinsic Gaussian distribution 
$\mathcal{N}(\mu,\sigma^{2})$ and a Gaussian error distribution 
$\mathcal{N}(0,e_{i}^{2})$ due to heteroscedastic errors for each of the 
metallicity measurements. The likelihood for obtaining the metallicity values 
$x_{i}={[Fe/H]}_{i}$, is given by \citep{hasc12,ivez14}
\begin{align*}
\mathcal{L}\left(\mu,\sigma^{2}\right)=& \prod_{i=1}^{N} \frac{1}{\sqrt{2\pi\left(\sigma^{2}+e_{i}^{2}\right)}}\exp{\left[-\frac{\left(x_{i}-\mu\right)^{2}}{2\left(\sigma^{2}+e_{i}^{2}\right)}\right].} 
\end{align*}       
The log likelihood is given by 
\begin{align*}
\ell \enspace=& -\frac{n}{2}\ln{2\pi}-\frac{1}{2}\sum \ln{\left(\sigma^{2}+e_{i}^{2}\right)}-\sum \frac{\left(x_{i}-\mu\right)^{2}}{2\left(\sigma^{2}+e_{i}^{2}\right)} \\
\Rightarrow \ell \enspace =& \text{constant}-\frac{1}{2}\sum \ln{\left(\sigma^{2}+e_{i}^{2}\right)}-\sum \frac{\left(x_{i}-\mu\right)^{2}}{2\left(\sigma^{2}+e_{i}^{2}\right)}.   
\end{align*}
{\bf The maximization of the log likelihood yields the values of the mean 
metallicity $\mu=-1.87$~dex and an intrinsic width of the distribution 
$\sigma=0.18$~dex for the underlying MDF. Therefore we find the final mean 
metallicity value to be $[Fe/H]=-1.87\pm0.18$~dex. This value of the mean 
metallicity of the SMC is in excellent agreement with the value of 
$-1.85\pm 0.33~\text{dex}$ as found by \citet{skow16}. The normalized MDF of 
$3522$ RRab stars is shown in Fig.~\ref{mdf}. The red colour solid line shows 
the fitted MDF with the individual metallicity values convolved with the 
individual Gaussian uncertainties while the blue solid line depicts the 
intrinsic Gaussian distribution of the MDF. Applying the above procedure of 
maximum likelihood estimation in the case of mean magnitude and reddening we 
find the following mean values for the SMC: $\mu_{0}=18.909\pm 0.148$ mag which
corresponds to a distance of $D = 60.506\pm 4.126$ kpc and 
$E(B-V)=0.066\pm 0.036$ mag. Here the uncertainties represent the intrinsic 
spread rather than the standard deviation of the mean. The values of the 
distance to the SMC, $D=60.58$~kpc and $60.0$~kpc, recently obtained by 
\citet{jacy17} and \citet{mura17}, respectively, are in good agreement with the 
present study although the values in these two studies are based on entirely 
different empirical/theoretical relations and calibrations.}              
\section{Period-Color Relation for SMC RRab Stars}
\label{pc}
\begin{figure}
\begin{center}
\includegraphics[width=0.5\textwidth,keepaspectratio]{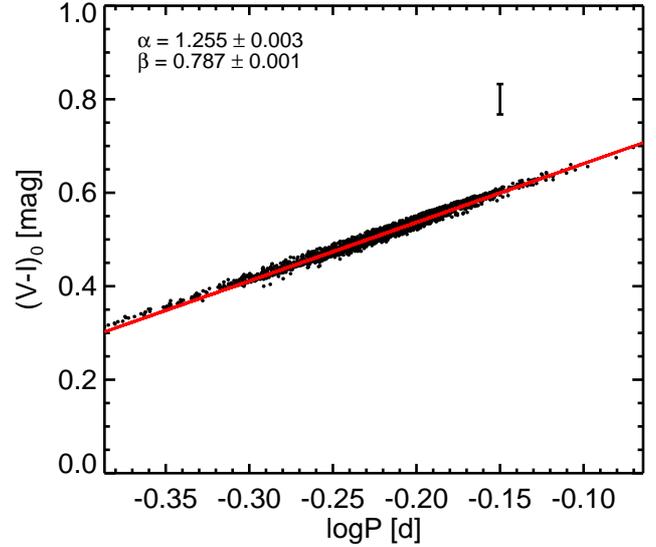}
\caption{{\bf Intrinsic period $(\log{P})$-colour ($(V-I)_{0}$) relation plot for 
 {\bf $3522$} SMC RRab stars modeled with a linear fit. The red solid line denotes the 
fitted line. The parameters of the fit are shown in the top left corner of the 
plot. Typical mean error bar of the measurements of $(V-I)_{0}$ is shown in 
the top right of the plot.}}
\label{pcolor}
\end{center}
\end{figure}

Statistically significant large number of SMC RRab stars with their reddening 
values  $E(V-I)$ determined in the present study provides a unique 
opportunity to explore the various possible relationships between the 
intrinsic colour $(V-I)_{0}$  and other available light curve parameters of 
these stars. We try to find out the empirical relationships between 
$(V-I)_{0}$ and various other parameters involving $P$, $A_{I},A_{V}$ and 
$[Fe/H]$. The various relationships of the following forms are tried:
\begin{align*}
(V-I)_{0}=& \alpha\log{P}+\beta \\
(V-I)_{0}=& \alpha\log{P}+\beta [Fe/H]+\gamma \\
(V-I)_{0}=& \alpha\log{P}+\beta A_{V}+\gamma \\
(V-I)_{0}=& \alpha \log{P}+\beta A_{I}+\gamma \\
\end{align*}  
In order to carry a regressional analysis using the above models, we use 
${\rm  lm()}$ function in R\footnote{http://www.r-project.org/} statistical
package. {\bf R} is an open source programming language and  a software 
environment for statistical computing and graphics. The results of the 
regressional analysis are shown in Table~\ref{coltable}.
\begin{table*}
\begin{center}
\caption{{\bf Intrinsic colour $(V-I)_{0}$ relations as a function involving 
$\log{P},[Fe/H],A_{V}$ and $A_{I}$, respectively.}}
\label{coltable}
\begin{tabular}{lcccccccr} \\ \hline
Relation & $\alpha$ & $\beta$ & $\gamma$ & $\sigma_{std}$& & $F$-value & $p(F)$& $R^{2}$ \\ \hline       
$(V-I)_{0}= \alpha\log{P}+\beta$ &$1.255\pm0.003$ &$0.787\pm0.001$& $-$ & $0.108$ &&$1.953\times 10^{5}$& $0.00$ &$0.982$ \\
$(V-I)_{0}= \alpha\log{P}+\beta  [Fe/H]+\gamma$ & $1.312\pm0.000$ & $0.025\pm0.000$ & $0.847\pm0.000$ & $0.005$ & &$4.813\times 10^{7}$&$0.00$&$1.000$ \\
$(V-I)_{0}= \alpha\log{P}+\beta  A_{V}+\gamma$ &$1.218\pm0.004$&$-0.009\pm0.000$&$0.786\pm 0.001$&$0.104$&&$1.038\times 10^{5}$&$0.00$&$0.983$ \\
$(V-I)_{0}= \alpha\log{P}+\beta  A_{I}+\gamma$ &$1.225\pm0.004$&$-0.011\pm 0.001$&$0.786\pm 0.001$&$0.106$ && $1.016\times 10^{5}$&$0.00$&$0.983$ \\ \hline
\end{tabular}
\end{center}
\end{table*}
The results thus obtained can be utilised to study the significance of 
addition of each of the predictor variables on the response variable. The 
larger value of $F$ in all the regressional analysis result indicates 
that given response variable $(V-I)_{0}$ can be approximated with any of these 
four relations.{\bf  But the simplest relationship involving lesser complexity 
parameters is 
\begin{align}
(V-I)_0=&1.255\log{P}+0.787,~~\sigma_{\text{std}}=0.108,
\label{percolor}
\end{align}     
where $\sigma_{\text{std}}$ represents the residual standard error per degrees 
of freedom. 
Fig.~\ref{pcolor} show the $(V-I)_{0}$ vs. $\log{P}$ plot of the $3522$ SMC 
RRab stars. The red solid line denotes the linear fit of the data points 
obtained from the regressional analysis. }

We now test this derived colour relation $\log{P}-(V-I)_{0}$ to find out 
the mean reddening to the Large Magellanic Cloud (LMC) using the RRab stars 
taken from the OGLE-IV database. For this, we download the suitable RRab 
file containing the information about the period ($P$), mean magnitudes in the 
$(V,I)$-band $(\overline{m}_{V},\overline{m}_{I})$ from the LMC OGLE-IV RRab 
database. We found $27138$ RRab stars which have mean magnitudes in both the 
two bands. The observed colour $(V-I)$ is then calculated from the given mean 
magnitude information.  Then using the period ($P$) information taken from the 
OGLE-IV database, the intrinsic colour $(V-I)_{0}$ for each of the RRab stars 
were obtained using equation~(\ref{percolor}). Making use of these 
information, the reddening value $E(V-I)$ for each of the $27138$ LMC RRab 
stars were determined. {\bf  A three-parameter Gaussian function fitted to the 
reddening distribution of the $27138$ yields the mean reddening of the LMC as 
$E(V-I)_{\text{LMC}}=0.096\pm0.067$ mag. The mean value of the reddening 
$E(V-I)_{\text{LMC}}=0.09 \pm 0.07$ mag of the LMC found by \citet{hasc11} 
obtained using completely different methods and calibrations is quite 
consistent with that obtained in this paper.}  The error in the above reddening 
estimation represents the actual reddening scatter and the observational 
error. The mean reddening to the LMC as obtained in this paper using the 
$\log{P}-(V-I)_{0}$ relation derived with the help of $3522$ SMC RRab stars 
is also comparable to $E(B-V)_{\text{LMC}}=0.08\pm0.04$ mag or 
$E(V-I)_{\text{LMC}}=0.10\pm0.05$  as quoted in \citet{cald91}. Therefore we 
find that the reddening determination of a galaxy can be made possible from 
calibrations of the present multicolour photometry of a statistically large 
number of OGLE RRab stars in the $(V,I)$-bands. Nonetheless we caution the 
reader that this method may not provide the accurate result for an individual 
star.
\section{Three dimensional structure of the SMC}
\label{morphology}
Larger areal coverage and availability of more number of RRab stars in the 
OGLE-IV phase  as compared to the data release of OGLE-III project as well as 
other earlier OGLE-projects provide a vital means to get an insight into the 
current understanding 
of the detailed three dimensional structure of the SMC. This will also 
facilitate in the refinement of various structure-related parameters of the 
SMC. Apart from that, the analysis of the distance determination and reddening 
estimation of each of the SMC RRab stars using their simultaneous light curve 
data available in $(V,I)$-band also provided an unbiased estimate in their 
determinations.                     
\label{structure}
We use the following steps leading to the parameter determinations of the 
three dimensional structure of the SMC \citep{deb15}:     
\begin{enumerate}
\item The right ascension ($\alpha$), the declination ($\delta$) and the 
distance ($D$) for each of the RRab stars obtained in the present study  are
converted into the corresponding Cartesian coordinates $(x,y,z)$. The 
$(x,y,z)$ coordinates are obtained using the transformation equations
\citep{vand101,wein01,deb15}:
\begin{eqnarray*}
\label{proj}
x=-D\sin(\alpha-\alpha_{0})\cos{\delta}, \\
y=D\sin{\delta}\cos{\delta_{0}}-D\sin{\delta_{0}}\cos{(\alpha-\alpha_{0})}\cos{\delta}, \\
z=D_{0}-D\sin{\delta}\sin{\delta_{0}}-D\cos{\delta_{0}}\cos{\alpha-\alpha_{0}}\cos{\delta}.
\end{eqnarray*}
\begin{figure}
\begin{center}
\includegraphics[width=0.5\textwidth,keepaspectratio]{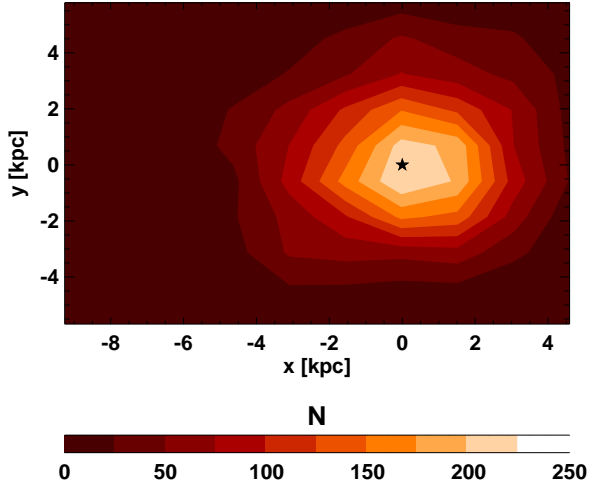}
\caption{Two-dimensional density contours of the SMC RRab stars in the present
study. The location of the centre of the SMC is shown as a star symbol.}
\label{contour}
\end{center}
\end{figure}
The two-dimensional density contours of the $3522$ SMC RRab stars in the 
present study is shown in Fig.~\ref{contour}.

The coordinate system of the SMC disk $(x^{\prime},y^{\prime},z^{\prime})$ is 
the same as the orthogonal system $(x,y,z)$, except that it is rotated around 
the $z$-axis by the position angle $\theta$ counterclockwise and around the 
new $x$-axis by the inclination angle $i$ clockwise. 
The coordinate transformations are \citep{vand101,wein01,deb15}:
\begin{align}
\label{eq:rot}
\begin{bmatrix} x^{\prime} \\ y^{\prime} \\ z^{\prime} \end{bmatrix} 
=  \begin{bmatrix} 
\cos{\theta} & \sin{\theta} & 0  \\
-\sin{\theta}\cos{i} & \cos{\theta}\cos{i} & -\sin{i} \\
-\sin{\theta}\sin{i} & \cos{\theta}\sin{i} & \cos{i} 
\end{bmatrix}\begin{bmatrix} x\\ y\\ z \end{bmatrix}
\end{align}
\item The Cartesian coordinate system $(x,y,z)$ has the origin in the centre 
of the SMC at $(\alpha,\delta,D)=(\alpha_{0},\delta_{0},D_{0})$. Here  
 we assume that the $z$ axis is pointed towards the observer and $x$-axis lies 
antiparallel to the $\alpha$-axis. The $y$-axis is taken parallel to the 
$\delta$-axis.	$D_{0}$ is the distance between the centre of the SMC and the 
observer. $D$ is the observer-source distance. $(\alpha_{0},\delta_{0})$ are 
the equatorial coordinates of the centre of the SMC. We take the centre of the 
SMC in the present study as $(\alpha_{0},\delta_{0}) = (0^{\rm h}53^{\rm m}31^{\rm s},-72^{\circ}59^{\prime}15^{\prime\prime}.7)$ \citep{subr12,deb15}. 

\item The errors in each of the Cartesian coordinates ($\sigma_{x},\sigma_{y},\sigma_{z}$) are obtained using the propagation of errors formula \citep{bevi03}. 

\item The observed distribution of the SMC RRab was modeled by a triaxial 
ellipsoid. Properties of the ellipsoid are obtained following the principal 
axes transformation method as described in \citet{deb14}.

\item The axes ratios $\frac{S_{i}}{S_{0}},\text{where}~i=0,1,2$, inclination 
of the longest axis along the line of sight $(i)$, position angle of line of nodes $(\theta_{lon})$ along with their associated errors were calculated using 
the Monte Carlo simulations carried out for $10^{5}$ steps  as discussed in 
\citet{deb15}.
\item The normalized distribution functions  having $10^{5}$ iterations of 
Monte Carlo simulations involving various SMC geometric parameters is  
obtained after binning with a proper binsize. The normalized distributions of 
the geometric parameters were found to approximate a Gaussian profile. Three 
parameter Gaussian profile fitting applied to the each of the distributions 
yields their mean and $\sigma$ values which are taken the as the true values 
and errors in these parameters.   
\end{enumerate} 
The normalized distributions of various structural parameters of the SMC 
obtained following the above steps in the present analysis using $3522$ RRab 
stars are shown in Fig.~\ref{structure}. The legend in each of the panels 
represents the mean and standard deviations of the distributions of the 
parameters obtained from thre three-parameter Gaussian fits which we quote as 
the geometrical values of the parameters  of the SMC.{\bf The following values 
of the parameters are obtained for the SMC with axes ratios $1.000\pm0.001,1.113\pm 0.002, 2.986\pm0.023$ and viewing angle parameters such as 
$i=3^{\circ}.156\pm0^{\circ}.188$ and $\theta_{\text{lon}}=38^{\circ}.027\pm0.577$. It should be noted that the position angle $(\theta)$ defined in equation 
(\ref{eq:rot}) is measured counterclockwise from the positive x-axis, i.e., 
from the west direction. The values of the position angles quoted as in this 
paper are given according to this direction.  However, in astronomical 
convention, position angles are always measured from the north ($0^{\circ}$) 
towards east ($90^{\circ}$). Therefore if measured from north, the position 
angle of line of nodes will be given by 
$\theta_{\text{lon}}=128^{\circ}.027\pm 0^{\circ}.577$. Also since the 
position angle is a line, its value can differ by an angle of $180^{\circ}$.}   

From the results obtained using the principal axis transformation method 
along with the Monte Carlo method for error estimation we find the lengths of 
the semi-major, semi-minor and intermediate axes as:  $S_{0}=12.229\pm 0.090$ 
kpc, $S_{1}=4.558\pm0.007$ kpc and $S_{2}=4.095\pm 0.004$ kpc, where  
$S_{0}>S_{1}>S_{2}$ \citep{deb14}. Following the above results we find that 
the longest axis viz. the $z$-axis is inclined by $3^{\circ}.156$ from the 
line of sight, i.e. the line of sight is almost along the $z$-axis of
the SMC.                
\begin{figure*}
\vspace{0.02\linewidth}
\begin{tabular}{ccc}
\vspace{+0.01\linewidth}
  \resizebox{0.32\linewidth}{!}{\includegraphics*{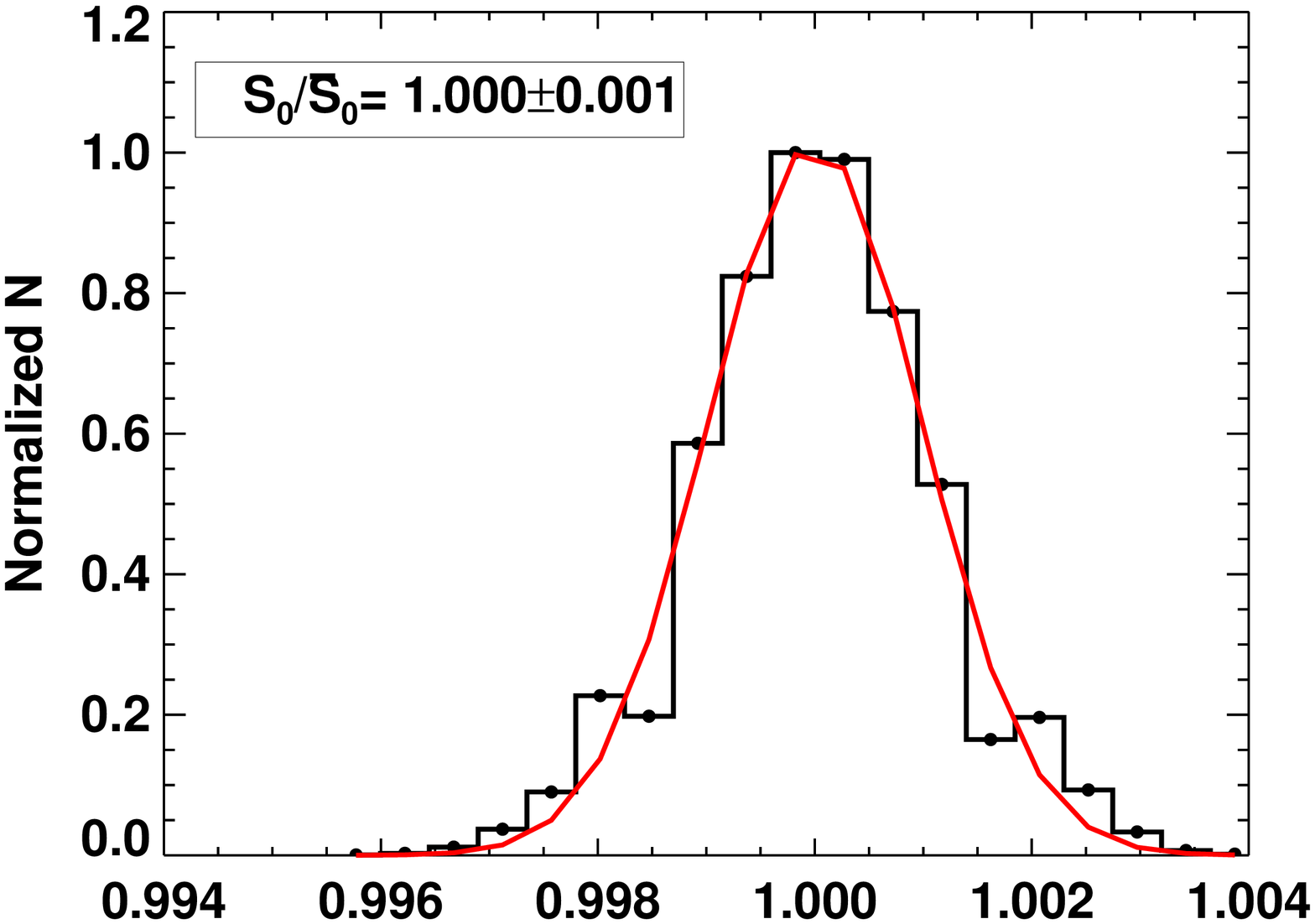}}&
   \resizebox{0.32\linewidth}{!}{\includegraphics*{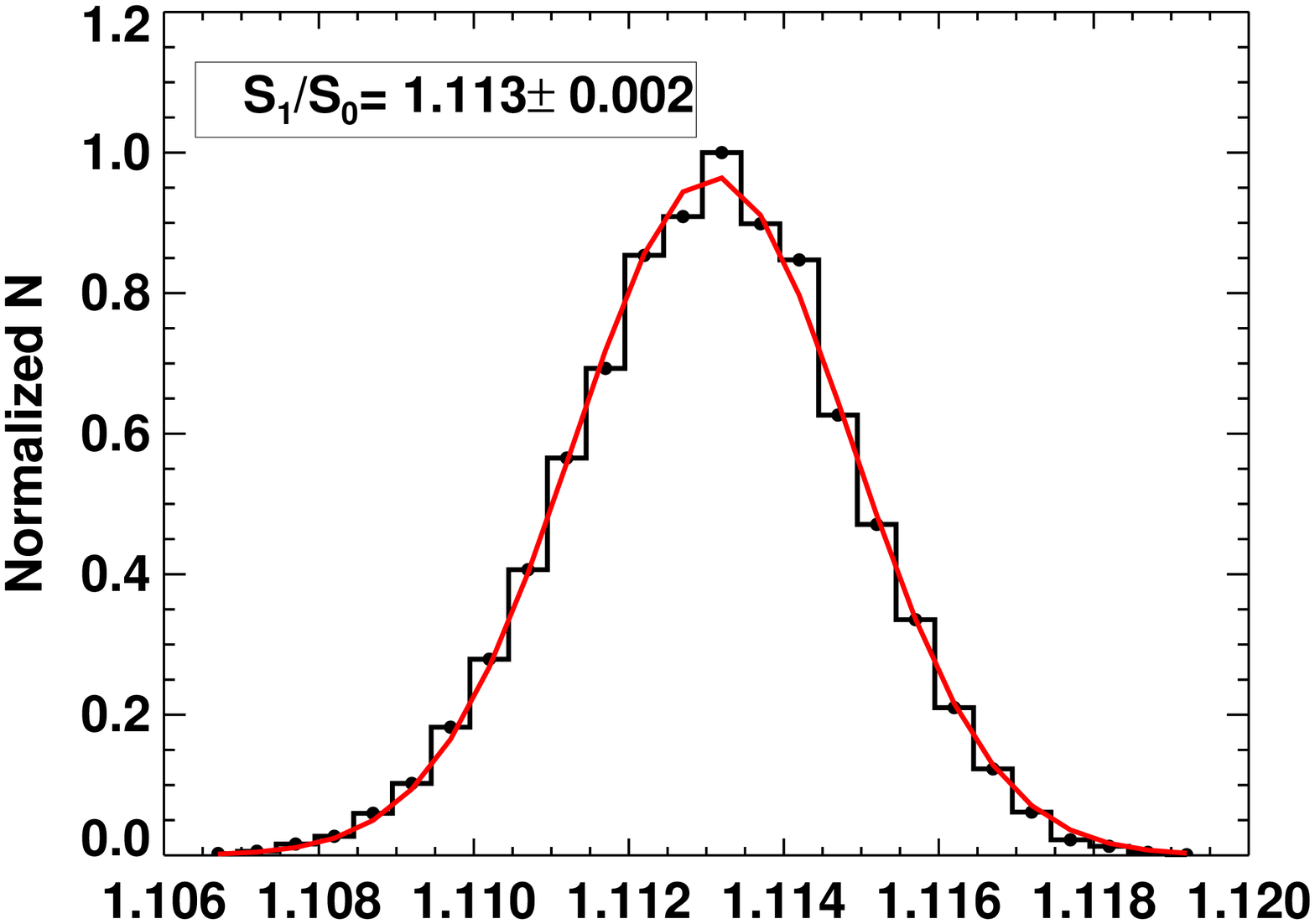}}&
  \resizebox{0.32\linewidth}{!}{\includegraphics*{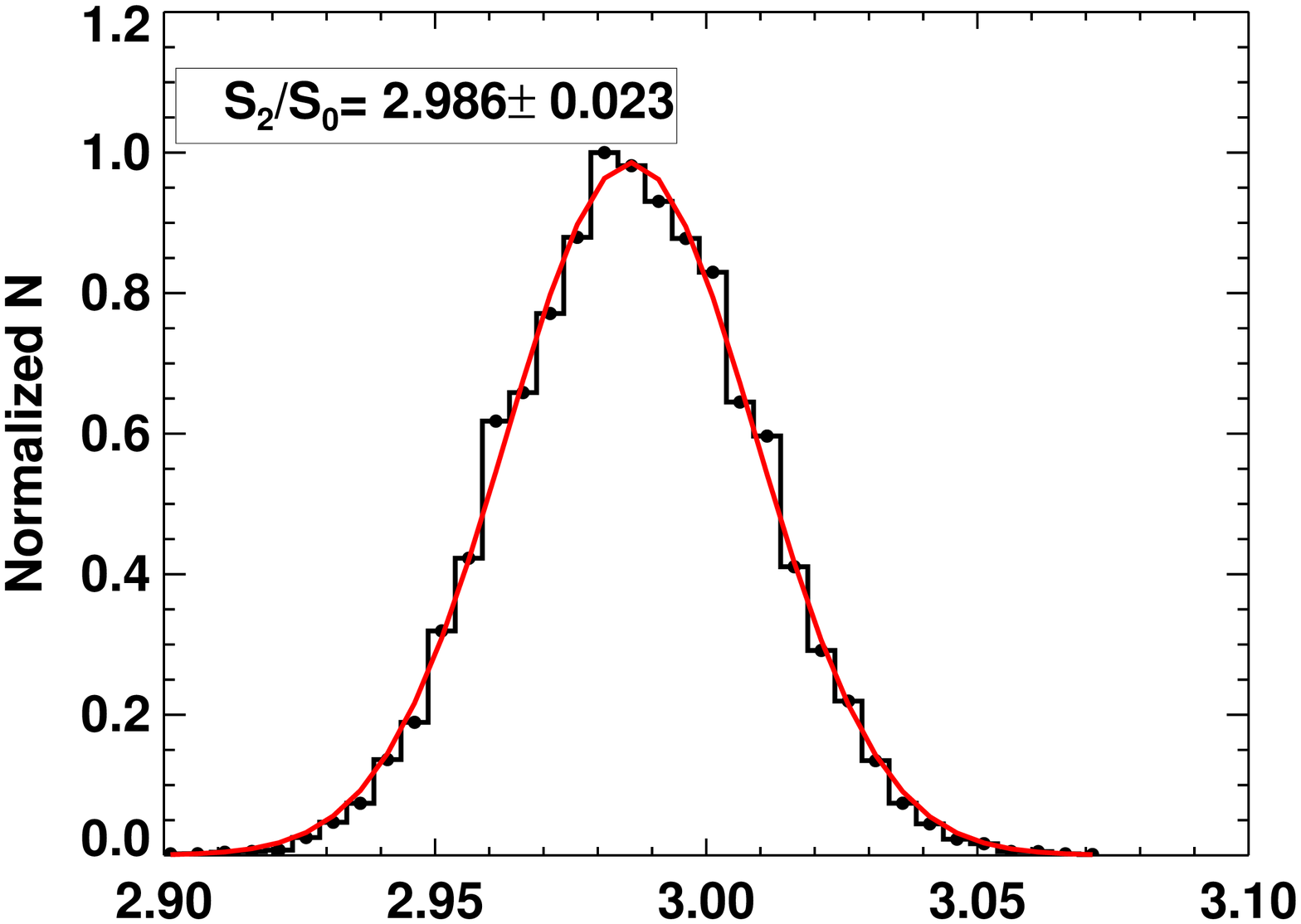}}\\  
\vspace{+0.01\linewidth}
\resizebox{0.32\linewidth}{!}{\includegraphics*{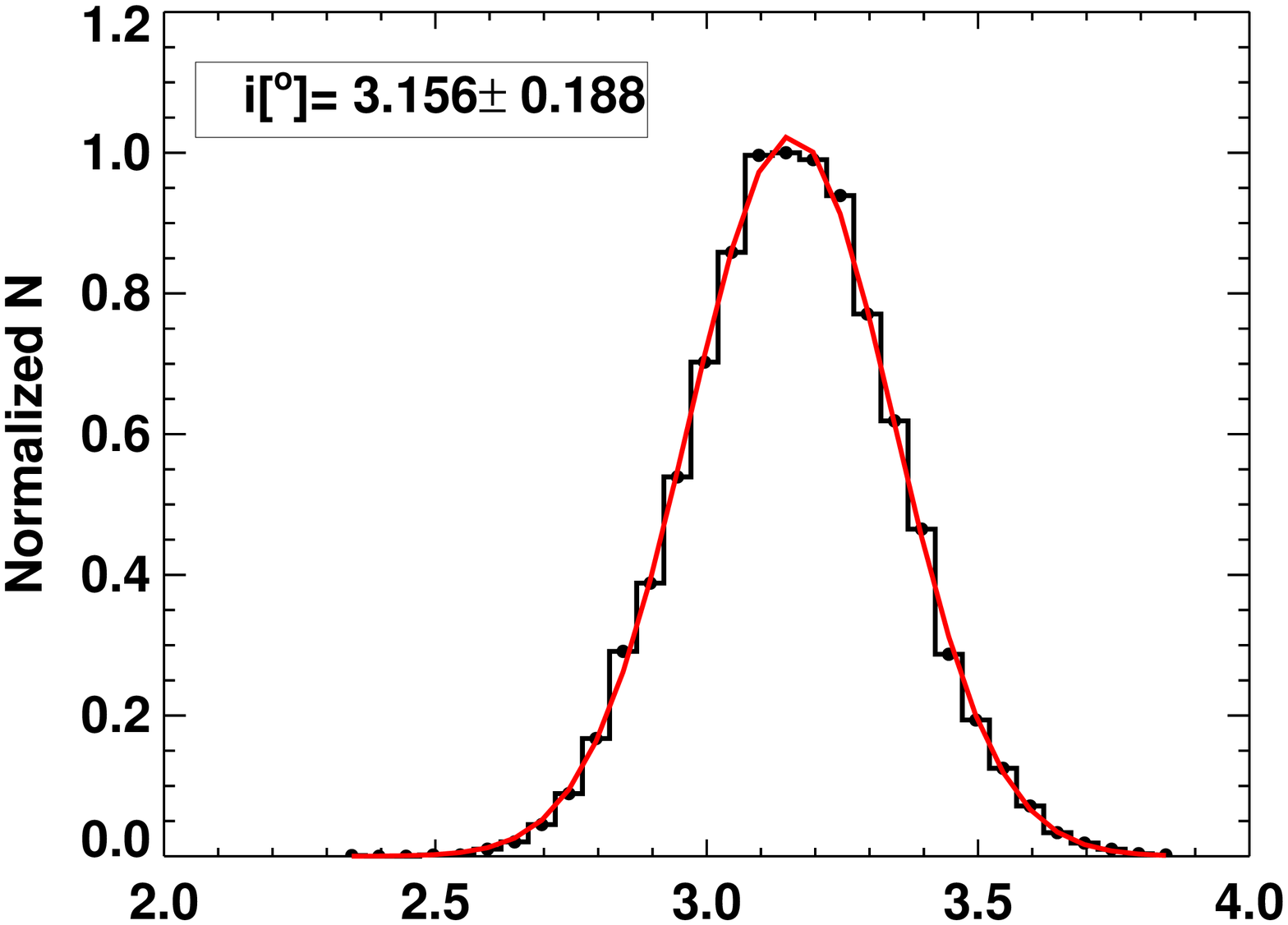}}&
  \resizebox{0.32\linewidth}{!}{\includegraphics*{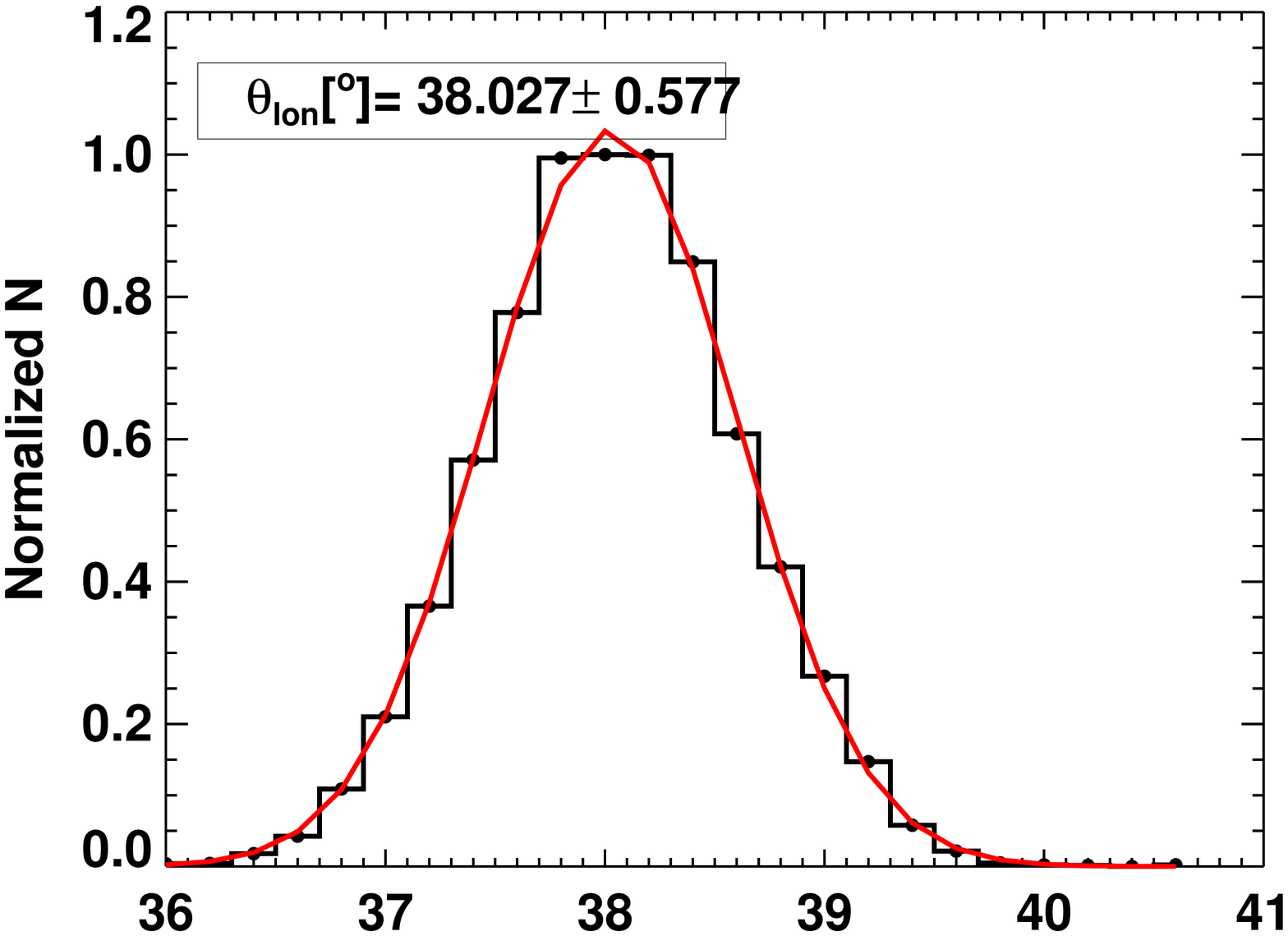}}\\
\vspace{-0.04\linewidth}
\end{tabular}
\caption{{\bf Histogram plots of the normalized distributions of various structural 
parameters of the SMC obtained using the Monte Carlo simulations of $10^{5}$ 
iterations applied on the principal axis transformation method.}}
\label{structure}
\end{figure*}

We also applied the simple plane-fitting procedure on the observed three 
dimensional distribution of the RRab star in Cartesian coordinates $(x,y,z)$.
The viewing angle parameters such as inclination ($i$) and position angle of 
line of nodes ($\theta_{\text{lon}}$) are obtained from a plane-fitting 
procedure of the form \citep{niko04,deb15}
\begin{equation}
\label{planefit}
z_{i}=c+ax_{i}+by_{i},~i=1,2,\dots,N,
\end{equation}                                         
where $N$ denotes  the number of data points. 
The inclination angle ($i$) can be obtained from the modeled 
parameters $(a,b,c)$ as  
\begin{displaymath}                         
i=\arccos{\left(\frac{1}{\sqrt{(1+a^2+b^2)}}\right)},
\end{displaymath}
Let us now define $\gamma = \arctan(|a|/|b|)$. Then position angle,
$\theta_{\text{lon}}$ can be obtained using \citet{deb15}
\begin{align*}
\theta_{\text{lon}} &=
\begin{cases}
\gamma & \text{if}~~a < 0~~\text{and}~~b > 0, \\
\gamma & \text{if}~~a > 0~~\text{and}~~b < 0,\\
\frac{\upi}{2} + \gamma & \text{if}~~a < 0~~\text{and}~~ b < 0, \\
\frac{\upi}{2} + \gamma & \text{if}~~a > 0~~\text{and}~~ b > 0, \\
0 & \text{if}~~a = 0~~\text{and}~~b~~\neq 0, \\
\text{sign}(a)\frac{\upi}{2} &~~\text{if}~~a \neq 0~~\text{and}~~b = 0,\\
\text{undef.} &~~\text{if}~~a = b = 0.
\end{cases}
\end{align*}

\begin{figure}
\begin{center}
\includegraphics[width=0.32\textwidth,keepaspectratio,angle=-90]{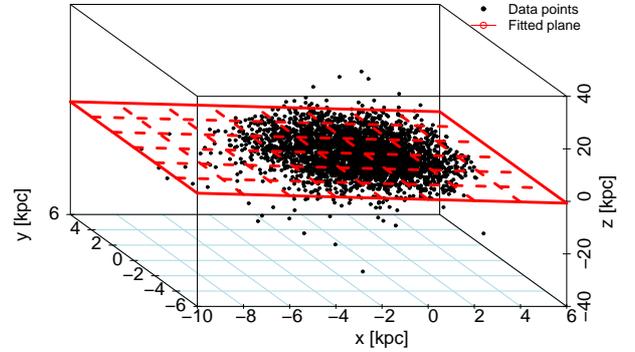}
\end{center}
\caption{{\bf Three dimensional $(x,y,z)$ distribution of the {\bf $3522$} SMC RRab stars.
The fitted plane with the parameters obtained from Bayesian MCMC plane fitting 
is also over plotted in red solid colours.}}
\label{fitting}
\end{figure}

{\bf We now apply a weighted plane-fitting procedure using the {\small{mpfitfunc}} 
in IDL in order to fit the three dimensional plane of the SMC 
\citep{mark09,mark12}. The fitting procedure yields the following values 
of the parameters: $c=0.890\pm 0.041,~a=-0.313\pm0.020,b=-0.207\pm 0.021$.
This gives the value of $i=20^{\circ}.580\pm0^{\circ}.625$ and 
$\theta_{\text{lon}} = 56^{\circ}.581\pm 3^{\circ}.159$. As measured from the 
north, the value of $\theta_{\text{lon}}$ will be 
$146^{\circ}.581\pm 3^{\circ}.159$. In order to validate the results obtained 
using the IDL routine  {\small{mpfitfunc}}, we further develop 
a Bayesian parameter estimation method in IDL to fit the three dimensional 
plane of the SMC. The parameter estimation in this case consists of three 
parts:  a plane fitting model, a likelihood function of the data and a prior 
distribution over the parameters. We have chosen uniform priors for the 
initial set of parameters. The posterior distribution of the sampling 
of the model parameters are obtained using  the Markov Chain Monte Carlo 
(MCMC) called the Metropolis-Hasting algorithm \citep{metr53,greg05}. The mean 
and standard deviations of the posterior distribution of the model parameters 
are treated as the best fit model parameters and their associated 
uncertainties.    The MCMC iteration was run for $10^{5}$ steps and the 
posterior distribution of the sampling of model parameters are noted for each 
step. The mean and standard deviation of PDF of the posterior probabilities of 
these model parameters are obtained as: $c=0.858\pm 0.042, a=-0.317\pm 0.018, 
b=-0.205\pm 0.027$. These parameters yield the values of the viewing angle 
parameters as: $i=20^{\circ}.682\pm 0^{\circ}.649$ and 
$\theta_{\text{lon}}=57^{\circ}.110\pm 3^{\circ}.747$. If the value of 
$\theta_{\text{lon}}$ is measured from the north, then 
its value will be $\theta_{\text{lon}}=147^{\circ}.110\pm 3^{\circ}.747$.  
The fitted plane with the parameters obtained using the Bayesian MCMC 
analysis is overplotted in Fig.~\ref{fitting}. However, the values of the 
parameters obtained from the three dimensional plane fitting should be taken 
with caution as the $z$-distribution for the SMC RRab stars has a larger 
spread and does not actually resemble a plane-like structure. Furthermore 
using this kind of simple three dimensional plane fitting algorithm we cannot 
determine the other structural parameters such as the axes ratios of the 
galaxy.}                 
\section{Comparison of SMC parameters obtained using SMOLEC'S (2005) 
Metallicity Relation and with other studies}
\label{compare}
\begin{figure*}
\begin{center}
\includegraphics[width=1\textwidth,keepaspectratio]{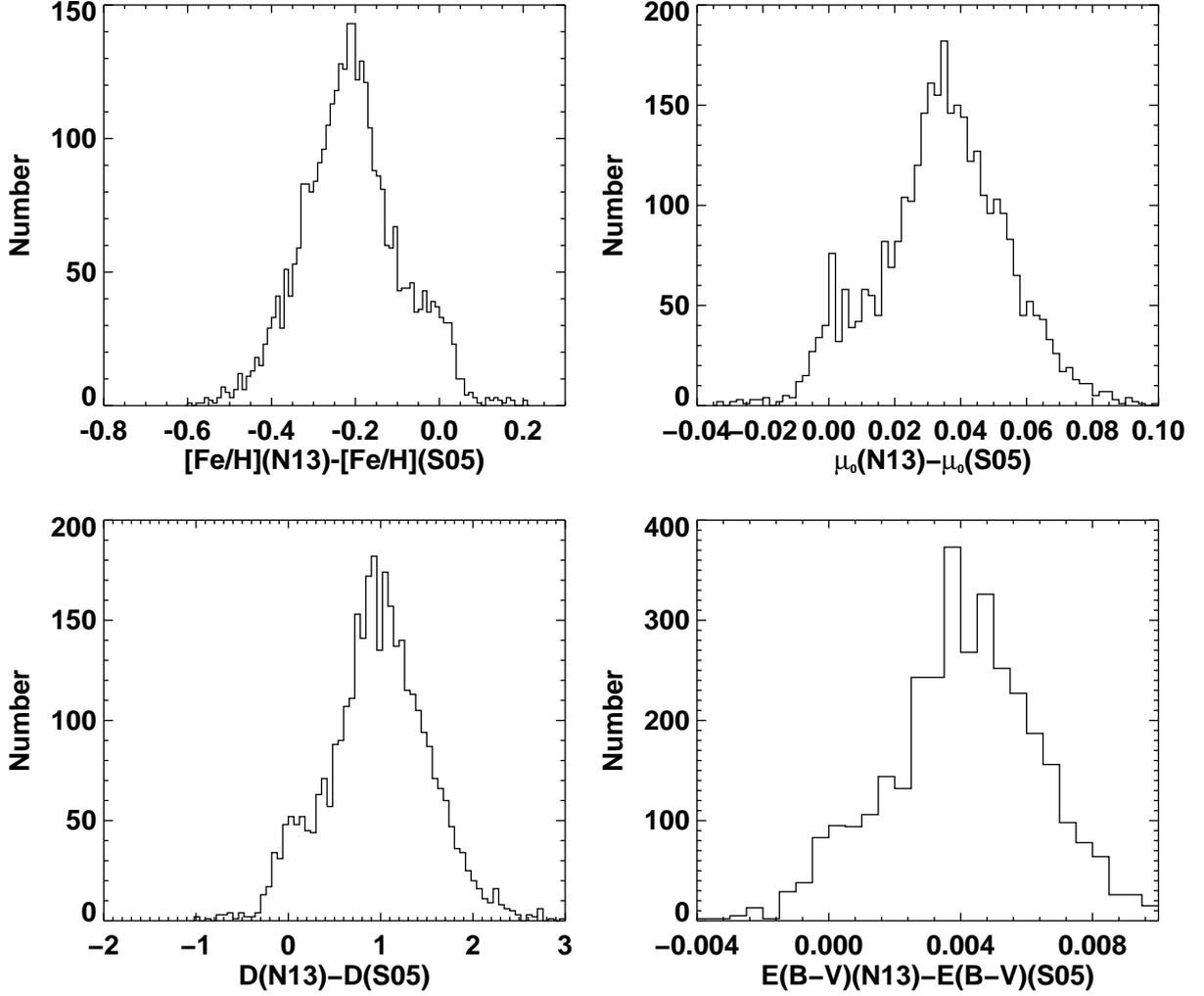}
\vspace{-0.1\linewidth}
\caption{{\bf Histogram plot of offsets for the $[Fe/H],~\mu_{0},~D$ and $E(B-V)$ 
values obtained using the \citet{neme13} (N13) and \citet{smol05} (S05) 
metallicity relations into the equations~(\ref{eq:absmag}) and (\ref{simul}). 
Mean differences of $\sim~-0.21$ dex, $0.04$ mag, $1$ kpc, $0.004$ mag
are obtained between the four parameters obtained using the N13 and S05 
relations.}}  
\label{hist}
\end{center}
\end{figure*}

{\bf
We now determine the mean distance and reddening of the SMC using 
equations~(\ref{simul}) and (\ref{eq:absmag}) with the $[Fe/H]$ values obtained 
from $[Fe/H]-P-\phi_{31}$ relation of \citep{smol05} which makes use  of the
$I$- band RRab data: 
\begin{align}
\label{smol05}
[Fe/H]_{JK}=& -3.142-4.902P+0.824\phi_{31},~~~\sigma_{\rm std}=0.18.  
\end{align}   
The above relation was based on the metallicity scale of JK95 metallicity. 
This linear relation was derived combining the light curve parameters of $28$
RRab stars with their complementary spectroscopic metallcities in the range
$-1.71~\text{dex}< [Fe/H]_{JK}< +0.01~\text{dex}$ \citep{smol05}. The relation 
given by equation~(\ref{smol05}) is transformed into the metallicity scale of 
ZW84 using the equation~(\ref{eq:zw84}). There are other studies, 
cf., \citet{hasc12,kais17} which make use of the following relation given by 
\citet{papa00} to convert $[Fe/H]_{JK}$ into the ZW84 scale:    
\begin{align*}
[Fe/H]=& 1.028[Fe/H]_{JK}-0.242. 
\end{align*}
But in a very detailed study by \citet{skow16}, it was demonstrated that the 
use of this relation gives the similar results as those of JK95 around 
$[Fe/H] \approx -1.4$ dex, but there exist large offsets of $-0.23$ dex 
at $[Fe/H] \approx -2.0$ dex and $0.53$ dex at $[Fe/H] \approx 0.0$ dex, 
respectively between the two scales. Also since there is no any clear 
derivation of how the \citet{papa00} relation was obtained \citep{skow16}, the 
use of \citet{papa00} relation is left out in the present study.

All the selection criteria of choosing a clean sample of RRab stars as 
discussed in Section~\ref{selection} remain the same except the criterion of 
metallicity which is taken here as $-2.50 \le [Fe/H]< 0$ dex. This reduces
the original $3931$ number of RRab stars to $3360$ for their further analysis. 
When the present $3360$ stars are matched with $3522$ stars obtained in 
Section~\ref{data} the number of common stars found in both are $3348$.
From the analysis of these $3348$ stars we have found the following mean 
values of the parameters of the SMC obtained using the \citet{smol05} 
metallicity relation: $[Fe/H]=-1.66\pm 0.13$ dex, $\mu_{0}=18.883\pm 0.149$ mag, $D=59.733 \pm 4.023$ kpc, $E(B-V)=0.062\pm 0.036$ mag. Here the uncertainties 
represent the spread of the population rather than the standard deviation of 
the mean. Making use of the \citet{smol05} metallicity relation the following 
values of the parameters are obtained for the SMC with axes ratios $1.000\pm0.001,1.109\pm 0.001, 2.791\pm0.018$ and viewing angle parameters such as: 
$i=3^{\circ}.791\pm0^{\circ}.155$ and 
$\theta_{\text{lon}}=38^{\circ}.779\pm0.442$. The following 
$\log{P}-(V-I)_{0}$ relation is obtained:
\begin{align*}
(V-I)_{0}=(1.272\pm 0.002)\log{P}+(0.797\pm 0.000).
\end{align*}
For this relation $\sigma_{\text{std}}=0.052$ denotes the 
residual standard error per degrees of freedom. This relation is almost 
identical to the relation given by equation~(\ref{percolor}) obtained using 
the \citet{neme13} metallicity relation. Histogram plots of offsets for the 
$[Fe/H],~\mu_{0},~D$ and $E(B-V)$ values obtained using the \citet{neme13} (N13) and \citet{smol05} (S05) metallicity relations, respectively into the 
equations~(\ref{eq:absmag}) and (\ref{simul}) are shown in Fig.~\ref{hist}. 
Mean systematic differences of $\sim~-0.21$ dex, $0.04$ mag, $1$ kpc, 
$0.004$ mag are obtained between the four parameters obtained using the 
\citet{neme13} and \citet{smol05} relations. The origin of these systematic 
differences are attributed to the systematic uncertainty in the $[Fe/H]$ 
values obtained using the \citet{smol05} relation as pointed out by 
\citet{skow16}.

The comparison between the distance-related parameters and structural 
parameters of the SMC obtained in the present study with their corresponding 
values found in the literature is shown in Table~\ref{comparison}. Although we 
find that the values of the SMC parameters obtained using the \citet{smol05} 
metallicity relation yield comparable values to those obtained using 
\citet{neme13} metallicity relation, there are subtle systematic biases 
present in the mean values of some of the parameters determined using the 
\citet{smol05} metallicity relation. The bias in the reddening value is almost 
negligible due to the presence of the expression $(\mu_{V}-\mu_{I})$ in the 
reddening estimation, which involves metallicities in each of the terms. 
Therefore any systematic bias present in metallicity in one of the terms is 
reduced/cancelled by the corresponding systematic bias in the other term. 
In fact we have found that the reddening map constructed based on the $[Fe]/H]$ 
relation of \citet{smol05} is quite similar to that obtained based on the 
\citet{neme13} relation. On the other hand systemtaic bias present in the 
\citet{smol05} metallicity relation does not get reduced/cancelled in the 
distance modulus calculation while using the second relation of 
equation~(\ref{simul}) and hence becomes significant. Due to the problem of 
systematic biases present in the \citet{smol05} metallicity relation towards 
the low and high metallicity ends, the mean value of the  metallicity obtained 
for the SMC and other parameters derived from it are quite unreliable using 
this relation \citep{skow16}. Although the consequences of these effects are 
reduced in a statistical analysis of a large population of RRab stars as in 
the present study they systematically effect the results on distance 
determinations using $M_{\lambda}-[Fe/H]$ relations. Since the calculation of 
metallicity using the \citet{neme13} metallicity relation is the most 
accurate, precise and free from any systematic bias we adopt the results 
obtained in the present study based on this relation as the final results.}   
\begin{table*}
\begin{center}
\caption{{\bf Comparison of distance-related parameters and structural 
parameters of the SMC obtained in the present study with their corresponding 
values available in the literature.}}
\label{comparison}
\scalebox{0.8}{
\begin{tabular}{|c|c|c|c|c|c|c|c|} \hline \hline \\
Reference & $\mu_{0}$ (mag) & $D$ (kpc) & $E(B-V)$ (mag) & $S_{1}/S_{0}$ & $S_{2}/S_{0}$ & $i[^{\circ}]$ & $\theta_{\text{lon}}[^{\circ}]$ \\ \hline
$1$ &-&-&$0.054\pm 0.029$&-&-&-&- \\
$2$ &$18.97\pm0.03(\rm stat.)\pm0.12(\rm sys.)$&-&-&-&-&-&- \\ 
$3$ &-&-&$0.056\pm0.048^{\dagger}$&-&-&-&- \\
$4$ & -& - & -  & $1.33$ & $6.47$ & $0^{\circ}.4$ & $74^{\circ}.4$ \\
$5$ &-&-&-& - &-&$7^{\circ}\pm15^{\circ}$ & $83^{\circ}\pm 21^{\circ}$ \\
$6$ &$18.93\pm 0.02$&$61.09\pm1.47$&-&-&-&-&- \\
$7$ &$18.965\pm0.025(\rm stat.)\pm 0.048(\rm sys.)$& $62.1\pm1.9$&-&-&-&-&- \\
$8$ &&&$0.048\pm0.039$&$1.310\pm0.029$&$8.269\pm0.934$&$2^{\circ}.265\pm 0^{\circ}.784$&$74^{\circ}.307\pm 0^{\circ}.509$ \\
$9$ &$18.96\pm 0.01$&$62.0\pm 0.3$&$0.071\pm0.004$&-&-&- \\
$10$ &-&-&-&$1.10$&$2.13$&&- \\ 
This work (N13) &$18.909\pm0.148$&$60.506\pm4.126$&$0.066\pm 0.036$& $1.113\pm0.002$ & $2.986\pm 0.023$ & $3^{\circ}.156\pm0^{\circ}.188$ & $38^{\circ}.027\pm 0^{\circ}.577$  \\
This work (S05) &$18.883\pm 0.149$&$59.733\pm 4.023$&$0.062\pm0.036$&$1.109\pm0.001$&$2.791\pm 0.018$&$3^{\circ}.791\pm0^{\circ}.155$&$38^{\circ}.779\pm 0^{\circ}.442$ \\ \hline
\end{tabular}
}
\end{center}
$1.~$\citet{cald86}; $2.~$\citet{szew09}; $3.~$\citet{hasc11}; $4.~$\citet{subr12}; $5.~$\citet{hasc12a}; $6.~$\citet{inno13}; $7.~$\citet{grac14}; $8.~$\citet{deb15}; $9.~$\citet{scow16}; $10.~$\citet{jacy17}. N13 - Using \citet{neme13} 
metallicity relation; S05 - Using \citet{smol05} metallicity relation; $^{\dagger}$ Converted into $E(B-V)$ using the relation $E(V-I)=1.26E(B-V)$. 
\end{table*}
\section{Summary and Conclusions}
\label{summary}
In this paper we have simultaneously utilised both the $V$- and $I$-band light 
curve data of more than $3000$ OGLE-IV SMC RRab stars in order to 
independently determine both the mean distance and reddening of the galaxy. 
We also study the three dimensional structure of the SMC using the distance 
distribution of each of the individual RRab stars along with their equatorial 
coordinates $(\alpha,\delta)$. The availability of a statistically large 
number of RRab light curve data simultaneously available in the 
multi-photometric $(V,I)$- bands with wider areal coverage being generated for 
this galaxy from the OGLE-IV photometric survey provides a unique opportunity 
to develop a more refined understanding of its distance, reddening and three 
dimensional structure. This newly obtained accurate and precise data have thus 
helped us in updating our recent knowledge about the distance, reddening and 
morphological structure of the galaxy. Based on the simultaneous analysis 
of $3522$ SMC RRab stars observed in two photometric bands $(V,I)$ the 
following results are obtained from the present study:
\begin{enumerate}
{\bf 
\item The true mean distance modulus $\mu_{0}$  and the mean reddening 
$E(V-I)$ for the SMC obtained from the light curve analysis of RRab stars are 
$18.909\pm 0.148$ mag  and $0.066\pm 0.036$ mag, respectively. The 
uncertainties quoted here represent the intrinsic spread in the population 
rather than the standard deviation of the mean. The mean distance to the SMC 
is obtained as $D = 60.505\pm 4.126$ kpc. We also find that the distance and 
reddening values obtained using the methodologies developed in this work are 
anticorrelated and is thus free from any possible systematic bias. 

\item One of the important results of our analysis is the reddening map of the 
SMC. From the reddening distribution of the SMC RRab stars the reddening map 
is constructed by computing the average reddening on a $10\times 10$ grid in 
$(x,y)$ coordinates. From the reddening map we find that the southern part of 
the SMC has relatively more reddening zones as compared to its northern part. \item The reddening values $E(V-I)$ obtained for each of the individual 
$3522$ RRab stars  along with their periods ($P$) taken from the OGLE-IV 
database have been utilised to derive a period ($P$)-colour ($(V-I)_{0}$) 
relation for these stars. The intrinsic colours are obtained from $(V-I)_{0}=(V-I)-E(V-I)$. The following PC relation was obtained:
\begin{align*}
(V-I)_{0}=(1.255\pm 0.003)\log{P}+(0.787\pm 0.001).
\end{align*}
For this relation $\sigma_{\text{std}}=0.108$ denotes the residual standard 
error per degrees of freedom. The above relation was tested on $27138$ 
OGLE-IV  LMC RRab stars to find the mean reddening to the galaxy as 
$E(V-I)=0.096\pm 0.067$ mag which is consistent with the $E(V-I)$ values for 
the LMC obtained using other tracers and different methodologies. This is a 
very useful and significant result on the ground that the above relation was 
obtained making use of various empirical relations available in the literature 
\citep{cate04,smol05,cate08,neme13,skow16} and this proves the robust validity 
of these relations in the application to a large database of RR Lyrae stars. 
The above relation will prove to be very useful in the estimation of mean 
reddening value of a host galaxy/globular cluster containing RRab stars quite 
easily.                                     
\item Approximating the three dimensional distribution  of the SMC RRab stars 
as ellipsoid, we have used the principal axes transformation method 
\citep{deb14,deb15} to find the axes ratios of the SMC: $1.000\pm0.001,1.113\pm 0.002, 2.986\pm0.023$ with $i=3^{\circ}.156\pm0^{\circ}.188$ and 
$\theta_{\text{lon}}=38^{\circ}.027\pm0.577$. These results are quite 
consistent with the axes ratios of $1,1.10, 2.13$ recently obtained by 
\citet{jacy17} using a triaxial ellipsoid fitting algorithm originally 
developed by \citet{turn99}. Their determinations are based on completely
different theoretical and empirical relations which are derived from entirely 
different calibrations. However the results obtained in this paper using the 
OGLE-IV dataset are somewhat different than those found by \citet{deb15} and 
\citet{subr12} using the entire data set of OGLE-III RRab stars. In the case 
of semi-major axis ratio $\frac{S_{2}}{\overline{S_{0}}}$, the difference is 
much more significant, the reason being attributed to the low areal coverage 
of the SMC obtained during  the OGLE-III Project. The results obtained using 
the the principal axis transformation method along with the Monte Carlo method 
for error estimation the lengths of the semi-major, semi-minor and intermediate 
axes are found as:  $S_{0}=12.229\pm 0.090$ kpc, $S_{1}=4.558\pm0.007$ kpc and 
$S_{2}=4.095\pm 0.004$ kpc, where  $S_{0}>S_{1}>S_{2}$ \citep{deb14}.     
} 
This is the first of a series devoted to the determination of the distance, 
reddening and deciphering the three dimensional structure of the SMC using the 
available simultaneous $(V,I)$-band RRab light curves. In the subsequent 
papers, we plan to study the distance, reddening and three dimensional 
structure of the SMC using the Type I and Type II classical Cepheids with the 
techniques and methodologies developed in this paper. The reddening maps 
produced independently using the classical Cepheids will provide an 
opportunity to compare and contrast the  reddening map of the SMC produced 
using the RRab stars in the present study.   
\end{enumerate}
\section*{Acknowledgments} 
The author thanks the OGLE-IV team for making their wealthy and invaluable 
variable star data publicly available for the welfare of the astronomical 
community. Thanks are  due to Science and Engineering Research Board (SERB), 
Department of Science \& Technology (DST), Govt. of India for financial 
support through a research grant D.O No. $\text{SB/FTP/PS-029/2013}$ under 
the  Fast Track Scheme for Young Scientists in Physical Sciences. The author 
would like to express his sincere gratitude to Prof. Dhruba J. Saikia, Cotton 
University for reading the first draft of this manuscript and providing many 
valuable comments and suggestions. {\bf The author acknowledges helpful 
discussions with  Abhijit Saha, Chow-Choong Ngeow and Shashi M. Kanbur 
while preparing the draft of the manuscript. Lastly, the author thanks 
the anonymous referee for making various helpful comments and useful 
suggestions which made the paper significantly relevant}. The use of 
arxiv.org/archive/astro-ph and NASA ADS databases is highly acknowledged. 
\bibliographystyle{mnras}
\bibliography{deb_bf}
\bsp
\label{lastpage}
\end{document}